\documentclass[aps,reprint,
superscriptaddress,
amsmath, amssymb,
floatfix]{revtex4-2}

\usepackage{graphicx}
\usepackage{subfigure}
\usepackage{dcolumn}
\usepackage{bm}

\usepackage{amsfonts}
\usepackage{xcolor}
\usepackage{array}
\usepackage{booktabs}
\usepackage{multirow}
\usepackage{mathrsfs}

\usepackage{hyperref}
\usepackage{cleveref}
\crefname{figure}{\textcolor{black}{FIG.}}{\textcolor{black}{FIG.}}
\crefname{table}{\textcolor{black}{Tab.}}{\textcolor{black}{Tab.}}
\crefname{equation}{\textcolor{black}{Eq.}}{\textcolor{black}{Eq.}}
\crefname{section}{\textcolor{black}{Section}}{\textcolor{black}{Section}}

\begin{document}
\title{Framework for Quasiperiodic Interfaces: Proximal Coincidence Point Set and Computation}  

\author{Suining Xiong}
\affiliation{School of Mathematical Sciences, Peking University, Beijing 100871, China}

\author{Wenwen Zou}
\affiliation{School of Mathematics and Computational Science, Hunan Key Laboratory for Computation and Simulation in Science and Engineering, Xiangtan University, Hunan 411105, China}

\author{Pingwen Zhang}
\email{pzhang@pku.edu.cn}
\affiliation{Institute for Math \& AI, Wuhan, Wuhan University, Hubei, 430072, China}
\affiliation{School of Mathematical Sciences, Peking University, Beijing 100871, China}

\author{Kai Jiang}
\email{kaijiang@xtu.edu.cn}
\affiliation{School of Mathematics and Computational Science, Hunan Key Laboratory for Computation and Simulation in Science and Engineering, Xiangtan University, Hunan 411105, China}

\begin{abstract}
We present a unified theoretical and computational framework that bridges mathematical quasiperiodicity with classical crystallographic models. Based on a rigorous cut-and-projection construction, the proposed proximal coincidence point set (PCPS) theory extends the classical coincidence site lattice model and further incorporates physically motivated perturbations encoding interfacial atomic mobility as well as visual indistinguishability. Spectral characteristics of PCPS naturally motivate a conserved Landau-Brazovskii model combined with projection method, yielding unified high accuracy in resolving quasiperiodic order across the entire interfacial plane. Representative quasiperiodic features are revealed in our numerical results, including generalized Fibonacci sequences in BCC $[110]$ tilt GBs, as well as repetitive patterns within the interstices of dislocation networks in low-angle BCC $[100]$ twist GBs and phase boundaries between BCC and face-centered cubic crystals. In high-angle BCC $[100]$ twist GBs, 12- and 8-fold quasicrystals emerge, while the PCPS theory combined with cyclotomic field projections further explains their restrictions of non-crystallographic symmetries. This framework not only provides a rigorous theoretical explanation for interface structures but also offers a path toward modeling other types of incommensurate structures.
\end{abstract}

\maketitle

\section{Introduction}
Interfaces are important mesoscale structures in crystalline materials.
Their atomic-scale configurations influence macroscopic properties, such as irradiation response, fracture resistance, and long-term durability \cite{li2015defects, chimi2001accumulation, ogata2005energy, xiao2016effect, lanccon2019incommensurate}.  
Interfaces are divided into grain boundaries (GBs) between identical phases and phase boundaries between different phases, together forming complex polycrystals \cite{watanabe2011grain, bollmann2012crystal}.  
Deciphering the atomic arrangement patterns and formation mechanisms of interfaces has long been a major research topic in materials science.

Early interface research focused primarily on periodic GB systems, establishing a series of mature theories and simulation methods \cite{friedel1926leccons, vitek1968intrinsic, vitek1980structure}.
coincidence site lattice (CSL) model, a classic theoretical framework for periodic GBs, describes the coincidence symmetry of lattices on either side of the GB, providing a basis for structural characterization of periodic GBs \cite{friedel1926leccons}.
Corresponding simulation methods such as molecular dynamics \cite{feng2015energy, seki2023incommensurate, frolov2018grain}, Monte Carlo methods \cite{banadaki2018efficient}, density functional theory \cite{lanccon2019incommensurate}, and phase field models \cite{yamanaka2017phase} have been successively established.
These methods assume periodic boundary conditions as a default to avoid surface effects, suitable for ideal systems with well-matched lattices and periodically repeating interface structures.

However, the complexity of interface structures far exceeds the scope of ideal periodic models.
As early as the 1980s, the experimental discovery of quasicrystals provided important insights into non-periodic interface research \cite{ShechtmanMet}.
Experimentally, quasiperiodic structural features observed in various alloy interfaces directly confirm that interfaces may exhibit quasiperiodic patterns \cite{shamsuzzoha1991atomic, deymier1991experimental, shamsuzzoha1999structural, shamsuzzoha2013geometrical, lanccon2019incommensurate, seki2023incommensurate}.  
In face-centered cubic (FCC) structures, $[100]$ $45^\circ$ twist, $[100]$ $45^\circ$ tilt, and $[110]$ $90^\circ$ tilt GBs all exhibit quasiperiodic structures \cite{shamsuzzoha1999structural, deymier1991experimental, dahmen2010correlation}.
Among these, $[110]$ $90^\circ$ tilt GB (with $(100)/(110)$ plane contact) contains 5-fold symmetric structural units \cite{penisson1998high, gautam2013atomic, bowers2016step}.  
In body-centered cubic (BCC) structures, $\Sigma 9$ GBs can form non-periodic densely packed icosahedral clusters \cite{seki2023incommensurate}.
These experimental phenomena cannot be reasonably explained by CSL model or traditional periodic interface theories.

Theoretically, researchers have established geometric descriptions of quasiperiodic interfaces, including structural unit model \cite{sutton1983structure, sutton1988irrational}, cut-and-project method \cite{rivier1988quasicrystals}, and high-dimensional projection theory \cite{gratias1988hidden, sutton1992irrational, sutton1989quasiperiodicity}, etc.
In the first two descriptions, quasiperiodic tilt GBs are confirmed to be a quasiperiodic arrangement of basic structural units and can also form a quasiperiodic sequence of alternating topological dislocations and isolated units through the cut-and-project method of a 2 dimensional lattice. By extension, quasiperiodic twist GBs are naturally hypothesized to be 2 dimensional quasicrystals, arising from planar cuts of higher-dimensional lattices with irrational orientations, analogous to the strip constructions used for tilt GBs.
In high-dimensional projection theory, interface structures are viewed as projections of 6 dimensional superlattices onto 3 dimensional physical space. 
The periodicity in high-dimensional space equivalently describes quasiperiodicity in low-dimensional space, enabling unified geometric characterization of general GBs and CSL GBs.
This approach can also be extended to analyze quasiperiodicity in heterogeneous interfaces.
Based on this, the generalized coincidence sites network (GCSN) was proposed \cite{GCSNAra}. 
However, as shown in \cref{sec:quasiperiodicity}, quasiperiodicity in general interfacial systems emerges only within the plane parallel to the interface, indicating that it is inherently a 2 dimensional structural characteristic rather than a fully 3 dimensional one.
A windowed modification of GCSN was subsequently introduced \cite{romeu2003interfaces} to address this issue by incorporating a finite window along the direction normal to the interface. While appropriate window choices can yield reasonable structural predictions, the selection process remains ad hoc and may complicate the description of the transition from bulk phases to the interfacial region. Moreover, these approaches are primarily geometric constructions and do not explicitly account for physical atomic mobility, limiting their direct integration with further numerical simulations.

In simulation, current techniques for quasiperiodic interfaces relying on periodic boundary conditions or free surfaces still have limitations \cite{li2019atomistic, lanccon2019incommensurate, penisson1998high, seki2023incommensurate}.
These simplifications not only introduce Diophantine approximation error \cite{jiang2025approximation} but also destroy the quasiperiodicity of interfaces under long-range observation, failing to accurately capture true interface structures.
Meanwhile, long-range observations require large-scale computational resources, which often exceed the capabilities of conventional simulation methods.
In summary, although the existence of quasiperiodic interfaces has been experimentally confirmed and related theoretical methods have made progress, achieving precise calculation and systematic characterization of quasiperiodic interfaces remains a major and unsolved challenge in materials science today.

In this work, we establish a unified framework for general interfacial systems, comprising a theoretical model together with an efficient numerical computation method. Within this framework, quasiperiodic interfacial structures arise naturally from the incompatibility between adjoining bulk phases and remain intrinsically confined to the interfacial region. This observation motivates our development of the proximal coincidence point set (PCPS), a mathematically well-posed model that elucidates the origin of quasiperiodic order and encodes it into the associated spectral structures. For numerical realization, we adopt the Landau–Brazovskii (LB) model \cite{brazovskii1975phase, zhang2008efficient} within the interfacial framework introduced in \cite{xu2017computing, jiang2022tilt, GBsoftware}. The projection method \cite{jiang2014numerical, jiang2018numerical, jiang2024numerical} is implemented on the Fourier–Bohr spectra derived from the theoretical PCPS construction. Spanning theoretical prediction to computational implementation, this framework provides an accurate and efficient tool for modeling arbitrary interfaces.

Our study focuses on three representative interfacial systems: BCC $[100]$ twist GBs, BCC $[110]$ tilt GBs, and BCC/FCC interfaces. Well-known structural features observed in previous studies \cite{feng2015energy, fu2020optical, schwartz1985atomic, schober1970quantitative} such as dislocation networks and periodic arrangements at special misorientations are recovered in our simulations. Results for BCC $[100]$ twist GBs further include the emergence of 8- and 12-fold interfacial quasicrystals as well as their symmetry selection within the PCPS framework. Beyond these cases, quasiperiodic structures arranged according to generalized Fibonacci sequences are observed in BCC $[110]$ tilt GBs, while BCC/FCC interfaces display low-angle interfacial patterns analogous to those observed in BCC $[100]$ twist systems. All cases clearly demonstrate representative quasiperiodic characteristics, including incommensurate spectra, finite local complexity (FLC), and repetitivity.

\section{Theoretical framework}

\subsection{Cut-and-projection scheme and quasiperiodicity}
\label{sec:quasiperiodicity}
Atomic configurations in crystalline materials can be modeled as point sets in Euclidean space $\mathbb{R}^3$. Restricting our discussion to an arbitrary point set $\Lambda$, the concepts of aperiodicity, quasiperiodicity, and quasicrystals have been rigorously defined and studied in foundamental mathematical works such as \cite{Baake2002, meyer1972algebraic, levine1984quasicrystals}. 

Mathematically, we call a set $\Lambda \subset \mathbb{R}^n$ \textit{aperiodic} if it satisfies the properties of finite local complexity (FLC) and has no periodic elements in its continuous hull $\mathbb{X}(\Lambda)$, see \cite{baake2013aperiodic}. The FLC property implies that, for any given compact set $K$, intersection $K \cap \{\Lambda + t\},\ t \in \mathbb{R}^n$, which could be regarded as an observation of $\Lambda$ through $K$, exhibits only finitely many distinct local patterns. On the other hand, the condition of non-periodicity in the hull $\mathbb{X}(\Lambda)$ can be intuitively understood as the inability to omit any defects in the structure by \textit{pushing them infinitely far away} through translation, thereby ensuring the intrinsic aperiodicity of the point set $\Lambda$.

\begin{figure}[!hpbt]
	\centering
	\includegraphics[width=0.45\textwidth]{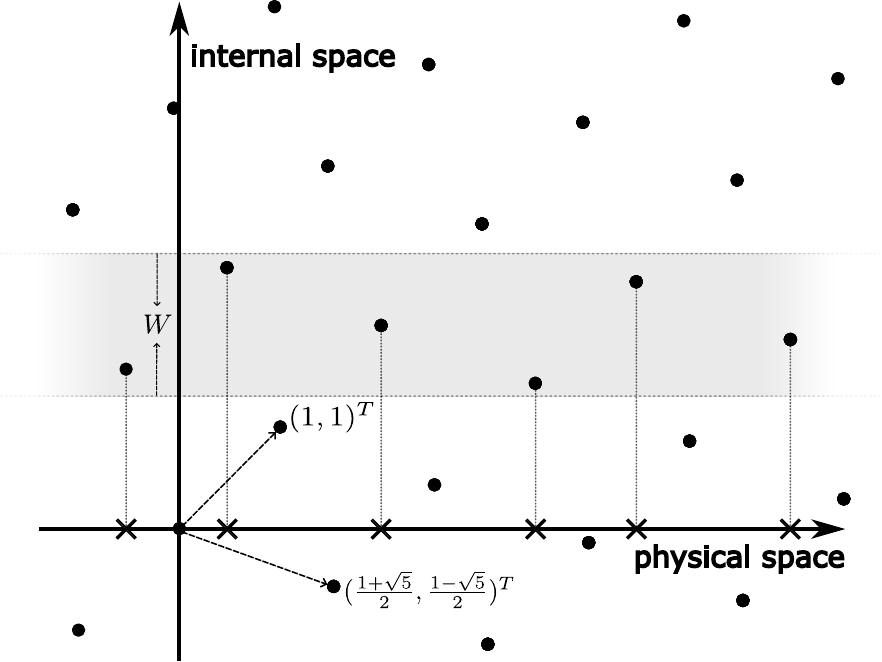}
	\caption{An example of CPS from $\mathbb{R}^2$ to physical space $\mathbb{R}$. The 2 dimensional lattice \(\Gamma\subset \mathbb{R}^2\) is generated by the basis vectors \((1,1)^T\) and \(\frac{1}{2}\left(1+\sqrt{5}, 1-\sqrt{5} \right)^T\). Cut window $W$ is selected as a close interval and $\pi_\perp^{-1}(W)$ is plotted in grey. The ``\( \times \)" marks the points in resulting cut-and-projection set \(\Lambda\).}
	\label{fig:CPSexample}
\end{figure}

Despite the definition of aperiodicity, there are three classical approaches to constructing aperiodic patterns: local matching rules, inflation and the cut-and-projection scheme (CPS) \cite{trevino2023aperiodic}. Among these, the CPS is one of the most widely used techniques. A general CPS construction can be described as follows: consider a $(m+n)$-dimensional Euclidean space $\mathbb{R}^{m+n}$, where the subspace $\mathbb{R}^n \subset \mathbb{R}^{m+n}$ is referred to as the \textit{physical space}, and its orthogonal complement $\mathbb{R}^m$ is called the \textit{internal space}. Since $\mathbb{R}^{m+n} = \mathbb{R}^m \times \mathbb{R}^n$, we can define the canonical projections from $\mathbb{R}^{m+n}$ onto $\mathbb{R}^m$ and $\mathbb{R}^n$ as follows
\begin{align*}
    &\pi_{\perp}: \mathbb{R}^{m+n} \longrightarrow \mathbb{R}^m, \quad (x,y) \mapsto x \in \mathbb{R}^m \\
    &\pi_{\parallel}\ :  \mathbb{R}^{m+n} \longrightarrow \mathbb{R}^n, \quad (x,y) \mapsto y \in \mathbb{R}^n.
\end{align*}
Let $\Gamma \subset \mathbb{R}^{m+n}$ be a lattice of rank $(m+n)$ satisfying the following conditions
\begin{itemize}\label{cpscondition}
    \item The restriction of $\pi_{\parallel}$ on $\Gamma$, namely $\pi_{\parallel}|_{\Gamma}$, is injective.
    \item $\pi_{\perp}(\Gamma)$ is dense in $\mathbb{R}^m$.
\end{itemize}
To construct a CPS set, we first choose a compact window $W \subset \mathbb{R}^m$ with non-empty interior. The lattice $\Gamma$ is then \textit{cut} through this window, yielding the subset $\pi_{\perp}^{-1}(W)\cap \Gamma \subset \Gamma$.
Then, we \textit{project} this subset onto the physical space using $\pi_{\parallel}$, resulting in the CPS set
\begin{align}
    \Lambda = \pi_{\parallel}\big( \pi_{\perp}^{-1}(W)\cap \Gamma \big).
\end{align}
\cref{fig:CPSexample} presents an example of CPS set.
Since patterns formed by CPS possess finitely generated Fourier-Bohr spectra, we refer to them as \textit{quasiperiodic}. Furthermore, if a quasiperiodic point set $\Lambda$ demonstrates a symmetry that is forbidden in crystallographic systems, such as $5$-, $8$-, or $12$-fold rotational symmetries, we call $\Lambda$ a \textit{quasicrystal}, which aligns with its definition in material science \cite{levine1984quasicrystals}.

A more general form of the CPS replaces the internal space $\mathbb{R}^m$ with a locally compact Abelian group $H$. In this framework, the point set $\Lambda$ generated by CPS are referred to as model set \cite{meyer1972algebraic}. Under certain regularity conditions model sets exhibit several desirable properties: model sets are Meyer sets and therefore Delone sets; model sets are also FLC sets and Repetitive, meaning that every finite patch (a finite subset of $\Lambda$, equivalently $K \cap \{\Lambda + t\}_{t \in \mathbb{R}^n}$) reappears infinitely many times in $\Lambda$ with a bounded upper distance between occurrences. Furthermore, translating the window by $\tau \in H$, yielding a shifted window $W' = \tau + W$, produces a different model set $\Lambda'$. Although the difference in window selections leads to  $\Lambda' \neq \Lambda$, CPS sets $\Lambda$ and $\Lambda'$ are still Locally Indistinguishable, namely every finite patch of $\Lambda$ can also be found in $\Lambda'$ and vice versa. This implies that it is impossible to distinguish  $\Lambda$ and $\Lambda'$ by visual inspection.

\subsection{PCPS model for interfaces}
\label{sec:CPS}
The interfacial system can be orthogonally decomposed into two directions: the $x$-axis, which is perpendicular to the interface, and the $y$-$z$ plane, which lies parallel to the interface. Along the $x$ direction, the two periodic bulk phases occupy the regions $x \to \pm \infty$ respectively, while the interface itself is confined to a narrow spatial interval. In contrast, within the $y$-$z$ plane, the atomic configuration at the interface is simultaneously influenced by both adjoining bulk phases across the entire plane. According to the discussion of quasiperiodicity, or more precisely, repetitivity in \cref{sec:quasiperiodicity}, quasiperiodic order can therefore only arise within the $y$-$z$ plane parallel to the interface.

\begin{figure}[!hpbt]
	\centering
	\subfigure[$\theta = 2\arctan(1/2)$]{
	    \includegraphics[width=0.22\textwidth]{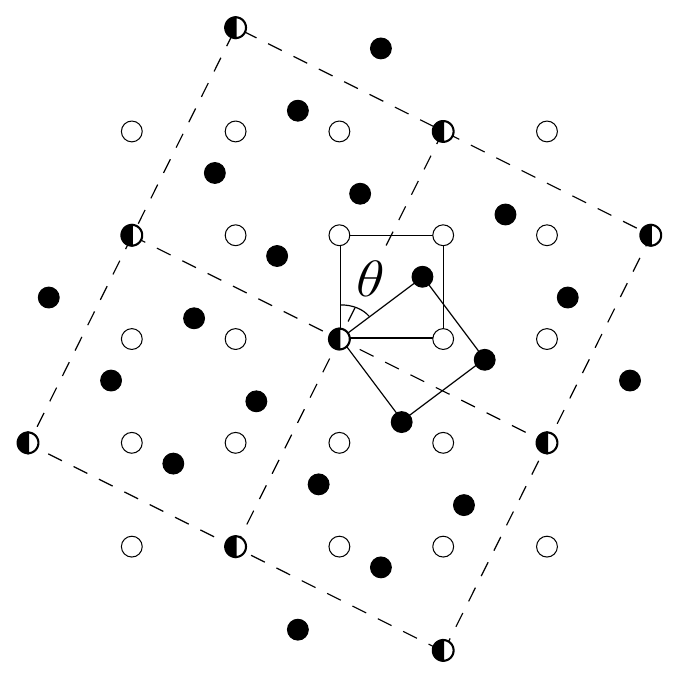}
	    \label{fig:DCP_1}}
	\subfigure[$\theta =30^\circ$]{
		\includegraphics[width=0.22\textwidth]{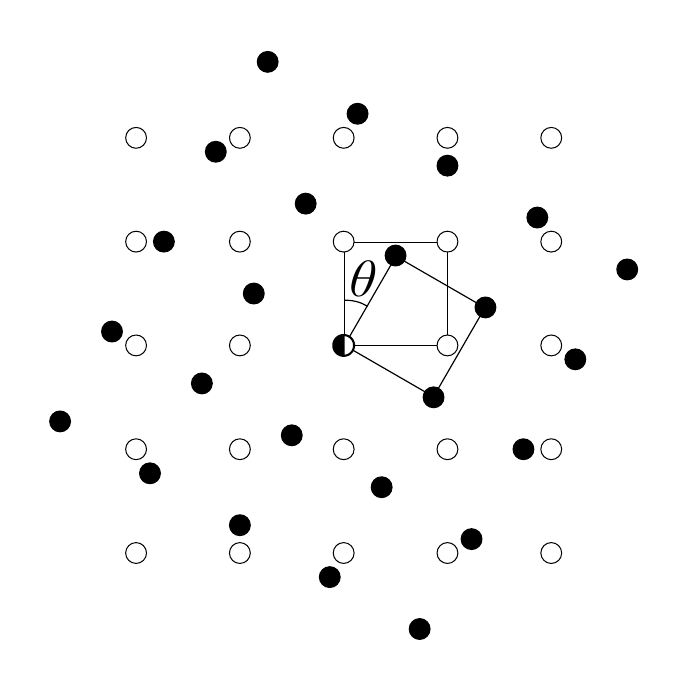}
		\label{fig:DCP_2}}
	\caption{Dichromatic pattern for two interpolating planar square lattice (a) $\theta = 2\arctan(1/2)$; (b) $\theta =30^\circ$. The misorientation between the black lattice and the white lattice is $\theta$. The dashed box in (a) shows the supercell for the CSL model.}
	\label{fig:DCP}
\end{figure}

Early theoretical efforts on interfacial structure modeling, particularly within the $y$-$z$ plane,  began with the CSL model \cite{ranganathan1966geometry, randle2024role}.
As shown in \cref{fig:DCP_1}, when misorientation $\theta$ satisfies $\tan (\theta/2) \in \mathbb{Q}$, the CSL model identifies the periodic coincidence sites formed between two adjacent planar square lattices, thereby enabling the prediction of interface structures. This approach has proven highly effective in characterizing the structural and symmetrical properties of interfaces, particularly GBs \cite{fu2020optical, frolov2018grain, wang2018grain} when the alignment of bulk lattices yields a commensurate and periodic configuration. However, for general cases, as illustrated in \cref{fig:DCP_2}, the assumption of strict periodicity no longer holds. Consequently, conventional models could hardly capture the quasiperiodicity of the complex patterns observed at the interface \cite{li2019atomistic, lanccon2019incommensurate}. 

To address these limitations, several generalized model have been developed \cite{levine1984quasicrystals, deymier1991experimental}, including the O-lattice theory \cite{bollmann2012crystal, zhang1993lattice}, the generalized coincidence sites network (GCSN) \cite{GCSNAra} and its subsequent windowed modification \cite{romeu2003interfaces}. Most of these generalization are founded on the concept of minimum strain points \cite{bollmann2012crystal, romeu2003interfaces}. In a local region near the interface, atoms from adjacent bulk phases interact more strongly when they are spatially close. Therefore, it becomes energetically favorable for interface atoms to occupy positions that minimize the local strain. These generalized models realize this principle by relaxing the strict coincidence condition of CSL into various forms of proximity conditions. 

In the GCSN framework, the two bulk lattices are idealized as planar lattices parallel to the interface but oriented differently, and atoms of two bulk phases are considered ``proximal" when they both lie within the intersection of their respective Voronoi cells. However, this construction fails when the interfacial projection of the bulk lattices does not yield a planar sublattice. In the subsequent windowed modification of GCSN, the same notion of proximity is retained but is applied directly to the original 3 dimensional bulk lattices, resulting in a quasiperiodic point set that fills the entire space $\mathbb{R}^3$. A small interval along the $x$-axis, referred to as a window, is then introduced to extract the transition from bulk phases to interface. Nevertheless, as discussed above, quasiperiodicity in interfacial systems arises only within the $y$-$z$ plane. The window selection in this framework lacks a consistent criterion, leading to the absence of a unified mathematical formulation. Furthermore, these models do not account for atomic mobility at the interface. The absence of physical flexibility and corresponding numerical frameworks may result in predictions that fail to capture representative structural features, such as rotational symmetries see \cref{sec:qc}.

Our PCPS model for general interfaces also serves as a natural generalization of classical CSL theory. To avoid the non-quasiperiodic behavior along the $x$-axis, we focus on the atomic configurations within the $y$-$z$ plane, where quasiperiodicity plays a dominant role. In our model, atoms of two bulk phases are considered ``proximal" if their midpoint lies near the interface and their mismatch (or distance) is small within a certain allowable error tolerance $\varepsilon$. Let $\Gamma_1$ and $\Gamma_2$ represent the bulk lattices in the two bulk phases occupying regions ${x>0}$ and ${x<0}$, respectively, with the interface located at ${x=0}$. Let $\bm{v}_i = (v_{ix},v_{iy},v_{iz})^T\in \Gamma_i,\ i=1,2$ be atoms of bulk phases. We treat $\bm{v}_1$ and $\bm{v}_2$ as $\varepsilon$-proximal pairs, if and only if
\begin{align}
    \Big|\bm{v}_1 - \bm{v}_2\Big|^2 + \Big|\frac{v_{1x}+v_{2x}}{2}\Big|^2 \leq \varepsilon^2.
\end{align}
Rigorously, the \textit{standard} PCPS $\Lambda$ for bulk phases $\Gamma_1$ and $\Gamma_2$ is defined as  
\begin{align}
    \Lambda = \left\{\frac{1}{2}\begin{pmatrix}
    v_{1y} + v_{2y}\\
    v_{1z} + v_{2z}
    \end{pmatrix} \ \Big| \ \bm{v}_1,\bm{v}_2 \text{ are } \varepsilon \text{-proximal pairs.}\right\}
\end{align}
Regarding that physically we have omitted the $\varepsilon$-size atomic misfit of atoms at the interface in our model, we therefore allow an arbitrary displacement of up to $\varepsilon/2$ for each point in standard PCPS $\Lambda$.
Consequently, any point set $\Lambda'$ within the closed $\varepsilon/2$-neighborhood of $\Lambda$ (under \textit{vague topology}, see \cite{baake2013aperiodic}) is also defined as a PCPS for bulk phases $\Gamma_1$ and $\Gamma_2$. This mobility assumption compensates for the missing structural information along the $x$-axis, ensuring physical realism.

Under our definition, the standard PCPS \(\Lambda\) for a given interface can be represented as a translation of the CPS set from \(\mathbb{R}^6\) to the physical space \(\mathbb{R}^2\), or as their unions. Consider the simplest case where \(\Gamma_1,\Gamma_2\) are given by  
\begin{align}
    \Gamma_1 &= \left\{ \bm{A} \bm{x} + \boldsymbol{\alpha} \ \Big|\ \bm{x} = (x_1,x_2,x_3)^T \in \mathbb{Z}^3 \right\},\\ 
    \Gamma_2 &= \left\{ \bm{B} \bm{y} + \boldsymbol{\beta} \ \Big|\ \bm{y} = (y_1,y_2,y_3)^T \in \mathbb{Z}^3 \right\}.
\end{align}
Here, \(\bm{A}=(\bm{a}_1,\bm{a}_2,\bm{a}_3)^T\) and \( \bm{B}=(\bm{b}_1,\bm{b}_2,\bm{b}_3)^T  \in \mathbb{R}^{3\times 3}\) represent the lattice vectors of \(\Gamma_1, \Gamma_2\), respectively, and \(\bm{\alpha}=(\alpha_1,\alpha_2,\alpha_3)^T, \boldsymbol{\beta}=(\beta_1,\beta_2,\beta_3)^T\) denote the relative position vectors of atoms within the unit cells. The process of identifying the $\varepsilon$-proximal pairs \(\bm{v}_1,\bm{v}_2\), i.e., the \textit{cut} process, is equivalent to finding the pairs \((\bm{x}^T, \bm{y}^T) \in \mathbb{Z}^6\) such that  
\begin{align}
\begin{aligned}
\bigg| 
\bigg( \begin{matrix}
\bm{A} & -\bm{B}
\end{matrix} \bigg)
\begin{pmatrix}
\bm{x} \\
\bm{y}
\end{pmatrix} &- \bigg(\boldsymbol{\beta}-\boldsymbol{\alpha} \bigg) \bigg| ^2\\ 
&+ \bigg|\frac{\bm{a}_1^T\bm{x}+\bm{b}_1^T\bm{y}+\alpha_1+\beta_1}{2}\bigg|^2 \leq \varepsilon^2.
\end{aligned}
\end{align}
The \textit{projection} process generating \(\Lambda\) is 
\begin{align}
\Lambda = \left\{
\dfrac{1}{2}
\begin{pmatrix}
\bm{a}_2^T &  \bm{b}_2^T \\ 
\bm{a}_3^T &  \bm{b}_3^T\end{pmatrix}
\begin{pmatrix}
\bm{x} \\
\bm{y}
\end{pmatrix}
+ 
\dfrac{1}{2} \begin{pmatrix}
\alpha_2 + \beta_2 \\ 
\alpha_3 + \beta_3
\end{pmatrix} 
\right\},
\end{align}
where \(\bm{x}\) and \(\bm{y}\) are selected during the \textit{cut} process. Summarizing these operations, we can define a six dimensional lattice \(\Gamma \subset \mathbb{R}^6\) as  
\begin{align}
\Gamma := \left\{ 
\left(\begin{array}{l}
\bm{A} \quad \ - \bm{B} \\
\bm{A}/2 \quad \ \bm{B}/2
\end{array}\right)
\begin{pmatrix}
\bm{x} \\
\bm{y}
\end{pmatrix} 
\ \Big| \ \begin{pmatrix}
\bm{x} \\
\bm{y}
\end{pmatrix}  \in \mathbb{Z}^6 
\right\}, \label{formula:CPS}
\end{align}
with a corresponding compact window \(W=\overline{B\left( \bm u, \varepsilon\right)} \subset \mathbb{R}^4\), where
\begin{align}
    \bm u = \begin{pmatrix}
    \bm \beta - \bm \alpha \\ 
    \frac{1}{2}(\alpha_1 + \beta_1)
    \end{pmatrix}\in \mathbb{R}^4.
\end{align}
The standard PCPS $\Lambda$ generated by our model is the CPS model set defined on lattice $\Gamma$ with cut window $W$, subject to an translation by $(\alpha_2 + \beta_2,\alpha_3 + \beta_3)^T/2$.

In general cases, \(\Gamma_1,\Gamma_2\) must be represented as  
\begin{align}
\begin{split}
\Gamma_1 &= \left\{ \bm{A} \bm{x} + \boldsymbol{\alpha} \ \Big|\ \bm{x} \in \mathbb{Z}^3,\ \boldsymbol{\alpha} \in F_1 \right\} = \bigcup_{\boldsymbol{\alpha} \in F_1} \Gamma_{1,\boldsymbol{\alpha}}, \\
\Gamma_2 &= \left\{ \bm{B} \bm{y} + \boldsymbol{\beta} \ \Big|\ \bm{y} \in \mathbb{Z}^3,\ \boldsymbol{\beta} \in F_2 \right\} = \bigcup_{\boldsymbol{\beta} \in F_2} \Gamma_{2,\boldsymbol{\beta}},
\end{split}
\end{align} 
where \(F_1,F_2\) are finite sets consisting of relative position vectors within unit cells. In these cases, \(\Lambda\) can be regarded as the union of point sets  
\begin{align}
    \Lambda = \bigcup_{\boldsymbol{\alpha}\in F_1, \boldsymbol{\beta} \in F_2} \Lambda_{\boldsymbol{\alpha}, \boldsymbol{\beta}}
\end{align}
in which \(\Lambda_{\alpha, \beta}\) represents the simpler cases generated by \(\Gamma_{1,\alpha}\) and \(\Gamma_{2,\beta}\) as discussed previously. All \(\Lambda_{\alpha,\beta}\) share the same 6 dimensional lattice $\Gamma$ regardless of their cut windows and translations. Therefore, general standard PCPS $\Lambda$ still maintain the properties of FLC, aperiodicity and quasiperiodicity.

\begin{figure}[!hpbt]
	\centering
	\includegraphics[width=0.4\textwidth]{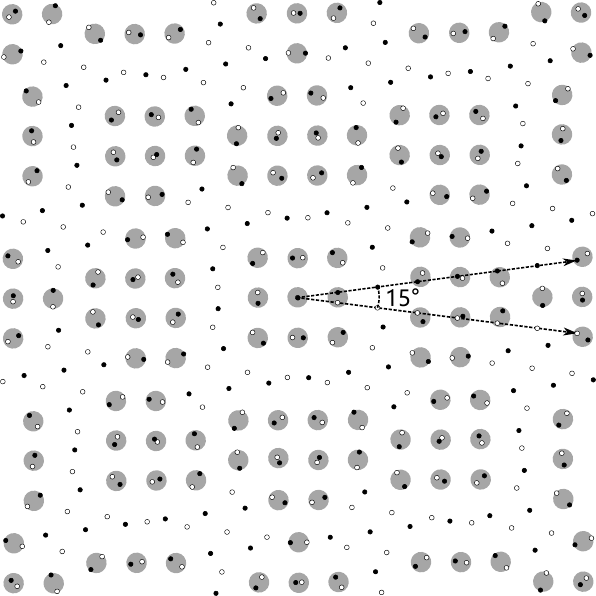}
	\caption{The PCPS model for BCC $[100]$ twist GB with twist angle \(\theta = 15^\circ\). Black and white lattices represent $\Gamma_{1,2}|_{GB}$.
	Large grey circles of diameter $\varepsilon$ are generated by black and white circles with distances less than $\varepsilon$, indicating the \(\varepsilon/2\) neighborhood within which each atom of vague PCPS should be located.}
	\label{fig:GBtheoretical}
\end{figure}

Since the PCPS is constructed as the union of several CPS sets and their translations, the choice of the parameter $\varepsilon$ directly corresponds with the point density $\rho$ of the PCPS $\Lambda \subset \mathbb{R}^2$ \cite{Baake2002}. In general, this relationship can be expressed as
\begin{align}
    \rho(\varepsilon) = 
    \frac{\pi^2 \varepsilon^4}{2|\det \bm{A}||\det \bm{B}|}\cdot\# F_1\cdot\# F_2 \label{formula:densrela}
\end{align}
where $\#F_1$ and $\#F_2$ denote the cardinalities of $F_1$ and $F_2$ respectively. \cref{fig:GBtheoretical} illustrates the PCPS model for two planar square lattices with misorientation $\theta=15^\circ$ and lattice constant $a$. The tolerance parameter is chosen as $\varepsilon\approx 0.451a$ to match the thresholded level set density of numerical results presented in \cref{fig:low_angle_qgb_15}.

\begin{figure}[!hpbt]
	\centering
	\includegraphics[width=0.4\textwidth]{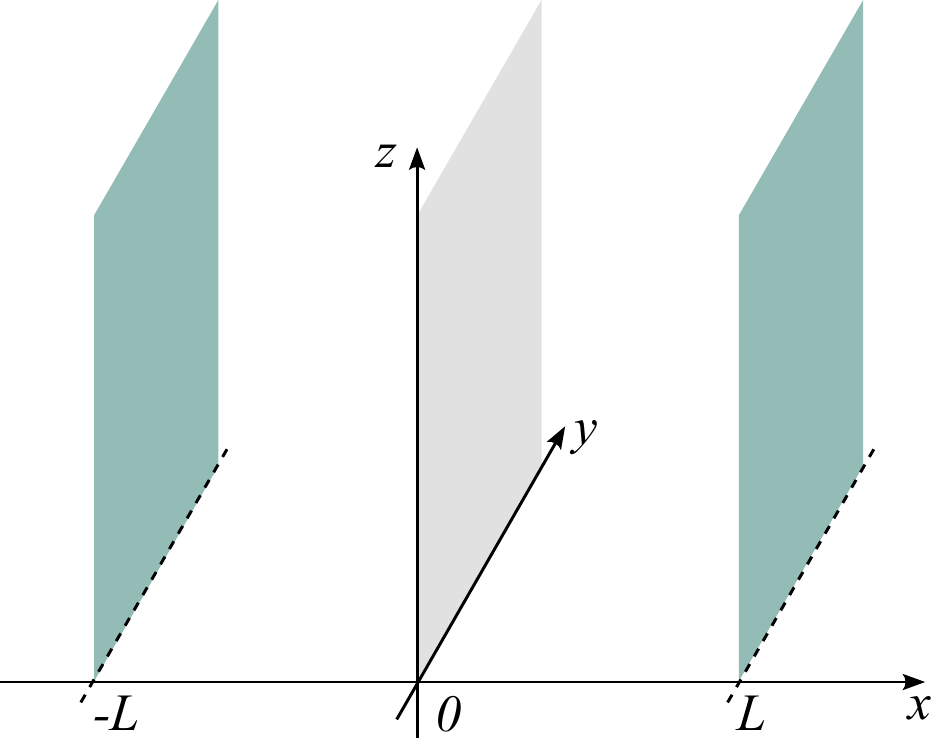}
	\caption{The space \(\mathbb{R}^3\) is divided into three regions. Interfaces between the bulk phases and the interfacial region located at \(x = \pm L\) are marked in green. The special plane at $\{x=0\}$ is highlighted in grey.}
	\label{fig:Numericalframe}
\end{figure}

\subsection{Conserved Landau-Brazovskii model for interfaces}
\label{sec:PFC}
Based on PCPS model developed in \cref{sec:CPS}, we establish a numerical framework for computing interfacial structures.
In our approach, the entire space $\mathbb{R}^3$ is partitioned into three regions, see \cref{fig:Numericalframe}.
The two bulk phases occupy the regions $x < -L$ and $x > L$, respectively, where $L$ is chosen sufficiently large to ensure complete structural relaxation within the intermediate transition region $-L \leq x \leq L$. The target interface is still positioned at $\{x=0\}$, in accordance with the PCPS model introduced above. Similar domain partition strategies can be found in \cite{xu2017computing, cao2021computing}.

Our computation model is based on conserved Landau-Brazovskii (LB) free energy functional, which was raised in 1975 and then proved to be effective in modeling phase transitions and crystalline structures, including both periodic and quasiperiodic systems \cite{brazovskii1975phase, zhang2008efficient, zou2025quasiperiodic}. A density distribution $\varphi(\bm{r}),\ \bm{r}=(x,y,z)^T$ is purposed to minimize interfacial LB free energy functional
\begin{align}
    \underset{\varphi}{\min} \underset{\Omega \to \mathbb{R}^2}{\lim} E[\varphi;\Omega], \label{general_minimization}
\end{align}
where $E[\varphi ; \Omega]$ is defined as 
\begin{align}
\begin{split}
    E[\varphi ; \Omega]=& \frac{1}{2L}\int_{-L}^L\frac{1}{V(\Omega)} \int_{\Omega}\left\{\frac{1}{2}[(\Delta+1) \varphi]^2\right.\\ 
    &+\left.\frac{\alpha}{2!} \varphi^2  
    -\frac{\gamma}{3!} \varphi^3+\frac{1}{4!} \varphi^4 \right\} \mathrm{d} \tilde{\bm{r}}\mathrm{d}x, \label{LB}    \\ 
    \tilde{\bm{r}} =& (y,z)^T,
\end{split} 
\end{align}
in the region $[-L,L]\times \Omega$ of volume $2L\cdot V(\Omega)$. The coefficients $\alpha$ and $\gamma$, associated with underlying physical parameters \cite{fredrickson1987fluctuation, leibler1980theory}, are selected to ensure the stability of the targeted interface system. Mass conservation condition imposes the constraint
\begin{align}
     \lim_{\Omega\to\mathbb{R}^2} \frac{1}{2L}\int_{-L}^L\frac{1}{V(\Omega)}\int_{\Omega}\varphi(x,\tilde{\bm{r}})\mathrm{d}\tilde{\bm{r}}\mathrm{d}x = 0
\end{align}
in the system. 

For different original bulk phases, periodicity allows the minimization problem to be restricted to a representative unit cell, with its LB free energy noted as \(E[\varphi]\). Accordingly, \( \varphi(\bm{r}) \) in periodic case can be expanded as a Fourier-Bohr series
\begin{align}
    \varphi(\bm{r}) = \sum_{\bm{k} \in \mathbb{Z}^3} \hat{\varphi}(\bm{k}) \exp\ [i(\bm Q \bm{k})^T \bm{r}],
\end{align}
where \( \{\bm Q\bm k\mid \bm k\in \mathbb{Z}^3\}\) denotes the reciprocal lattice of the crystal, and \( \hat{\varphi}(\bm{k}) \) are the Fourier-Bohr coefficients associated with frequency \( \bm{k} \). As illustrated in \cref{fig:Numericalframe}, the two bulk phase regions \( x > L \) and \( x <  -L \) are occupied by distinct bulk fields $\varphi_{\pm}(\bm r)$, each characterized by its own reciprocal lattice \(\{\bm Q_\pm \bm k\mid \bm k\in \mathbb{Z}^3\}\) where the wavevector matrices take the block decomposition
\begin{align}
    \bm Q_\pm =\begin{pmatrix}
    \bm q_\pm^T \\ 
    \bm P_\pm
    \end{pmatrix}. 
\end{align}
In order to accommodate the fourth-order derivatives in the LB free energy and the continuity conditions at $x = \pm L$, we separate the longitudinal $x$ and transverse $y$-$z$ directions and represent the bulk fields as
\begin{align}
    \varphi_{\pm}(x,\bm{k}) &= \sum_{\bm{k} \in \mathbb{Z}^3} \hat{\varphi}_{\pm}(x,\bm{k}) \exp\ [i(\bm P_\pm \bm{k})^T \tilde{\bm{r}}],\\ 
    \hat{\varphi}_{\pm}(x,\bm{k}) &= \hat{\varphi}_{\pm}(\bm{k})\exp\ [i (\bm q_\pm^T \bm k) x].
\end{align}
The interfacial field $\varphi(x,\tilde{\bm r})$ is then required to match the bulk solutions at $x=\pm L$ through Dirichlet-type boundary conditions
\begin{align}
\begin{split}
    \left.\frac{\partial^s \varphi(x, \tilde{\bm{r}})}{\partial x^s}\right|_{x=\pm L}=\left.\frac{\partial^s \varphi_{\pm}(x, \tilde{\bm{r}})}{\partial x^s}\right|_{x=\pm L}, ~s=0,1.
\end{split}
\end{align}

Different from periodic cases, the structure within the intermediate transition region is governed by contributions from both two bulk phases. As a result of PCPS model, quasiperiodicity naturally emerges at the $y$-$z$ plane of the interface. By analyzing its Fourier-Bohr properties, the intrinsic frequency spectrum is identified as 
\begin{align}
 \{\bm{P}\bm{k}\mid \bm{k}\in \mathbb{Z}^6\},\ \bm{P} = (\bm{P}_{+}, \bm{P}_{-}) \in \mathbb{R}^{2 \times 6}.
\end{align}
To numerically model the resulting interfacial structures in the entire domain, we utilize the projection method \cite{jiang2014numerical, jiang2018numerical}, specifically designed for quasiperiodic systems, to construct a Fourier-Bohr expansion of \( \varphi(x,\tilde{\bm{r}}) \) over $x\in [-L,L]$,
\begin{align}
    \varphi(x,\tilde{\bm{r}}) = \sum_{\bm{k} \in \mathbb{Z}^d} \hat{\varphi}(x, \bm{k}) \exp\left[i(\bm{P} \bm{k})^T \tilde{\bm{r}} \right].
\end{align}
The projection matrix $\bm{P}$, which corresponds to the Fourier-Bohr spectrum identified in our analysis above, encodes the embedding of the 6 dimensional periodic lattice into physical space. It is subsequently reduced such that its columns are linearly independent over $\mathbb{Q}$, without altering the spectrum $\{ \bm{P}\bm{k} \mid \bm{k} \in \mathbb{Z}^6 \}$; thus, $d \leq 6$. When the interface is commensurate (periodic), the projection dimension $d$ of $\bm P$ can be reduce to $d = 2$; otherwise, the interface is quasiperiodic with $3\leq d \leq 6$. In this representation, the unbounded physical interfacial $y$-$z$ plane is realized as a 2 dimensional irrational submanifold embedded in the compact $d$ dimensional torus $\mathbb{T}^d$. The quasiperiodic field $\varphi(x,\tilde{\bm r})$ is thereby implicitly represented as the pullback of a $d$ dimensional periodic function. While the physical coordinate $\tilde{\bm r}$ ranges over $\mathbb{R}^2$, all frequencies governing quasiperiodic order are encoded within the high dimensional torus via $\bm P$. As a result, the quasiperiodic field over the entire interfacial plane is uniquely determined by this embedding \cite{jiang2024numerical, fan2025representation}.
Shifting the numerical treatment of quasiperiodic interfaces from truncation of the unbounded physical domain to spectral discretization on the high dimensional torus $\mathbb{T}^d$, the projection method enables uniformly high accuracy in resolving quasiperiodic order across the entire interface, as discussed below.

\subsection{Discretization and optimization of LB model}
To numerically calculate the bulk and interfacial structures, discretizations on target free energy functional $E[\varphi]$ are imposed. For periodic bulk phases, \( E[\varphi] \) in \cref{LB} can be expressed in terms of Fourier-Bohr coefficients as
\begin{align}
\begin{aligned}
E[\varphi] =\; & \frac{1}{2} \sum_{\bm{k} \in \mathbb{Z}^3} \left[ \left(1 - |\bm{Q} \bm{k}|^2 \right)^2 + \alpha \right] \hat{\varphi}(\bm{k}) \hat{\varphi}(-\bm{k}) \\
& - \frac{\gamma}{3!} \sum_{\bm{k}_1 + \bm{k}_2 + \bm{k}_3 = \bm{0}} \hat{\varphi}(\bm{k}_1) \hat{\varphi}(\bm{k}_2) \hat{\varphi}(\bm{k}_3) \\
& + \frac{1}{4!} \sum_{\bm{k}_1 + \bm{k}_2 + \bm{k}_3 + \bm{k}_4 = \bm 0} \hat{\varphi}(\bm{k}_1) \hat{\varphi}(\bm{k}_2) \hat{\varphi}(\bm{k}_3) \hat{\varphi}(\bm{k}_4).
\end{aligned}
\end{align}
The functional derivative of the energy with respect to \( \hat{\varphi}(\bm{k}) \), which governs the optimization dynamics, takes the form
\begin{align}
\begin{aligned}
\frac{\partial E[\varphi]}{\partial \hat{\varphi}(\bm{k})} =\; & \left[ \left(1 - |\bm{Q} \bm{k}|^2 \right)^2 + \alpha \right] \hat{\varphi}(-\bm{k}) \\
& - \frac{\gamma}{2!}\sum_{\bm{k}_1 + \bm{k}_2 = - \bm{k}} \hat{\varphi}(\bm{k}_1) \hat{\varphi}(\bm{k}_2) \\
& + \frac{1}{3!}\sum_{\bm{k}_1 + \bm{k}_2 + \bm{k}_3 = -\bm{k}} \hat{\varphi}(\bm{k}_1) \hat{\varphi}(\bm{k}_2) \hat{\varphi}(\bm{k}_3).
\end{aligned}
\end{align}
This expression, when truncated to a finite number of modes, can be efficiently evaluated using the Fast Fourier Transform (FFT) algorithm \cite{zhang2008efficient, ShenJiebook}, thus enabling efficient optimization of bulk phases $\varphi(\bm{r})$.

In the interfacial region $x\in [-L,L]$, distinct discretization strategies are adopted along the \( x \)-direction and within the \( y \)-\( z \) plane. For the latter, where quasiperiodicity naturally emerges, the field \( \varphi(x, \tilde{\bm{r}}) \) is expanded via a Fourier-Bohr series, with a canonical spectral cut-off imposed by restricting the Fourier modes to \( \|\bm{k}\|_{\infty} \leq N_F \), thereby ensuring the FFT compatibility. Along the \( x \)-direction, where quasiperiodicity is absent, Dirichlet-type boundary conditions must be satisfied. These restrictions motivates the use of Generalized Jacobi polynomial basis \( \{J_{i+4}^{-2,-2}(x)\}_{i=0}^{N} \) for the expansion of \( \varphi(x, \tilde{\bm{r}}) \) \cite{ShenJiebook}. Numerical integration of LB free energy is then performed using the Legendre-Gauss quadrature rule, employing \( (2N+1) \) Legendre-Gauss points \( \{x_j\}_{j=0}^{2N} \subset [-L, L] \). This rule achieves an algebraic precision of degree $4N$, thereby enabling accurate numerical integration of the LB free energy. Consequently, the discretized field \( \varphi \) can be represented in terms of the degrees of freedom
\begin{align}
    \hat{\Phi} = \left\{ \hat{\varphi}(x_j, \bm{k}) \;\middle|\; \|\bm{k}\|_\infty \leq N_F,\ j = 0, \dots, 2N \right\},
\end{align}
leading to the approximation \( E[\hat{\Phi}] \).

 The discretized LB free energy \( E[\hat{\Phi}] \) naturally decomposes into a quadratic part and a non-quadratic convolutional part
\begin{align}
    E[\hat{\Phi}] = G[\hat{\Phi}] + F[\hat{\Phi}],
\end{align}
where the quadratic term \( G[\hat{\Phi}] \) takes the form
\begin{align}
    G[\hat{\Phi}] = \frac{1}{2 L} \sum_{j=0}^{2 N} \omega_j \sum_{\|\bm{k}\|_{\infty} \leq N_F} &\frac{1}{2} (1- |\bm{P} \bm{k}|^2+\partial_x^2 )^2 \nonumber\\
    &\hat{\phi}\left(x_j, \bm{k}\right) \hat{\phi}\left(x_j,-\bm{k}\right)
\end{align}
whose gradient $\nabla G[\Phi]$ and hessian $\nabla^2 G[\Phi]$ can be efficiently computed, and the nonlinear term \( F[\hat{\Phi}] \) is given by
\begin{align}
    F[\hat{\Phi}] = &\frac{1}{2 L} \sum_{j=0}^{2 N}\omega_j \Bigg[\frac{\alpha}{2}\sum_{\|\bm{k}\|_\infty \leq N_F} \hat{\varphi}(x_j, \bm{k}) \hat{\varphi}(x_j,-\bm{k})\nonumber\\ 
    &-\frac{\gamma}{3!} \sum_{\bm{k}_1+\bm{k}_2+\bm{k}_3=0} \hat{\varphi}\left(x_j, \bm{k}_1\right) \hat{\varphi}\left(x_j, \bm{k}_2\right) \hat{\varphi}\left(x_j, \bm{k}_3\right)\nonumber\\ 
    &+\frac{1}{4!} \sum_{\bm{k}_1+\bm{k}_2+\bm{k}_3+\bm{k}_4=0} \hat{\varphi}\left(x_j, \bm{k}_1\right) \hat{\varphi}\left(x_j, \bm{k}_2\right)\nonumber\\ &\quad\quad\quad\quad\quad\quad\quad\quad\quad\quad\hat{\varphi}\left(x_j, \bm{k}_3\right) \hat{\varphi}\left(x_j, \bm{k}_4\right)\Bigg].
\end{align}
Here, \( \{\omega_j\}_{j=0}^{2N} \) are the associated quadrature weights of integration with \( \{x_j\}_{j=0}^{2N} \) . $\nabla F[\hat{\Phi}]$ can also be computed efficiently using FFT.

During the energy minimization, the conservation of mass constraint is ultimately reduced to a 1 dimensional linear constraint on the discretized coefficient set $\hat{\Phi}$
\begin{align}
    e_\omega ^T\hat{\Phi} = \frac{1}{2L} \sum_{j=0}^{2 N} \omega_j \hat{\varphi}\left(x_j, \bm
{0}\right)= 0.
\end{align}
Following our decomposition of $E[\hat{\Phi}]$, we adopt an accelerated proximal gradient method for optimization \cite{jiang2020efficient, bao2024convergence}. Denote $\hat{\Phi}^{(l)}$ as the result of iteration $l$, we proceed our optimization as
\begin{align}
\hat{\Psi}^{(l)}&=\hat{\Phi}^{(l)}+w_l\left(\hat{\Phi}^{(l)}-\hat{\Phi}^{(l-1)}\right),\\
\hat{\Phi}^{(l+1)}&=\underset{e_{\omega}^T\hat{\Phi}=0}{\mathrm{argmin}} \left\{\beta_l G[\hat{\Phi}]+\right.\nonumber\\ \quad\quad\quad\quad\quad&\left.\frac{1}{2}\left\|\hat{\Phi}-\left(\hat{\Psi}^{(l)}-\beta_l \nabla F\left[\hat{\Psi}^{(l)}\right]\right)\right\|^2\right\},
\end{align}
where $\beta_l$ is standard Bregman step size, $w_l$ is the extrapolation weight, and \( \left\| \cdot \right\| \) is \( \ell^2\) norm.
The initial condition $\hat{\varphi}^{(0)}(x,\bm{k})$ is constructed by using a sigmoid-type function $\sigma(x)$ to smoothly connect bulk phases $\hat{\varphi}_{\pm}(\bm{r})$
\begin{align}
    &\sigma(x)=\frac{1+\tanh (0.1 x)}{2},\\
    \hat{\varphi}^{(0)}(x,\bm{k}) = (1&-\sigma(x))\hat{\varphi}_{-}(x,\bm{k}) + \sigma(x)\hat{\varphi}_{+}(x,\bm{k}).
\end{align}
This initial configuration also aligns with the viewpoint of the minimum strain points described in previous theoretical works. To enable sufficient relaxation of the interfacial structure, we select the interfacial computation width as $L=8 a$.

\begin{figure*}[!hpbt]
	\centering
	\includegraphics[width=0.04\textwidth]{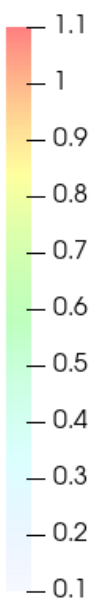}
	\subfigure[$\theta = 2\arctan(1/3) \approx 36.87^\circ$]{
	    \includegraphics[width=0.3\textwidth]{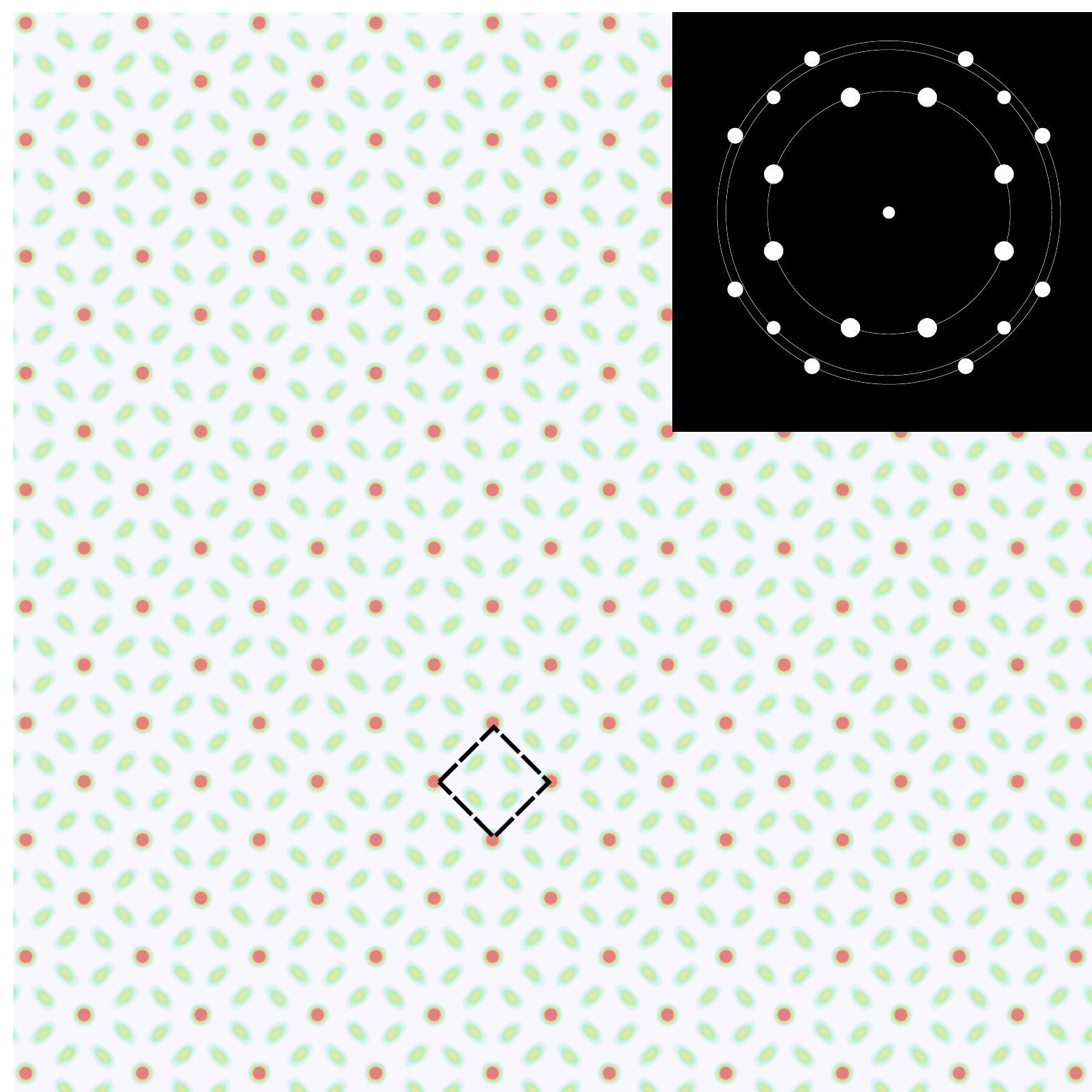}}
	\subfigure[$\theta = 2\arctan(1/9) \approx 12.68^\circ$. Thresholded level set and atomic configuration.]{
		\includegraphics[width=0.3\textwidth]{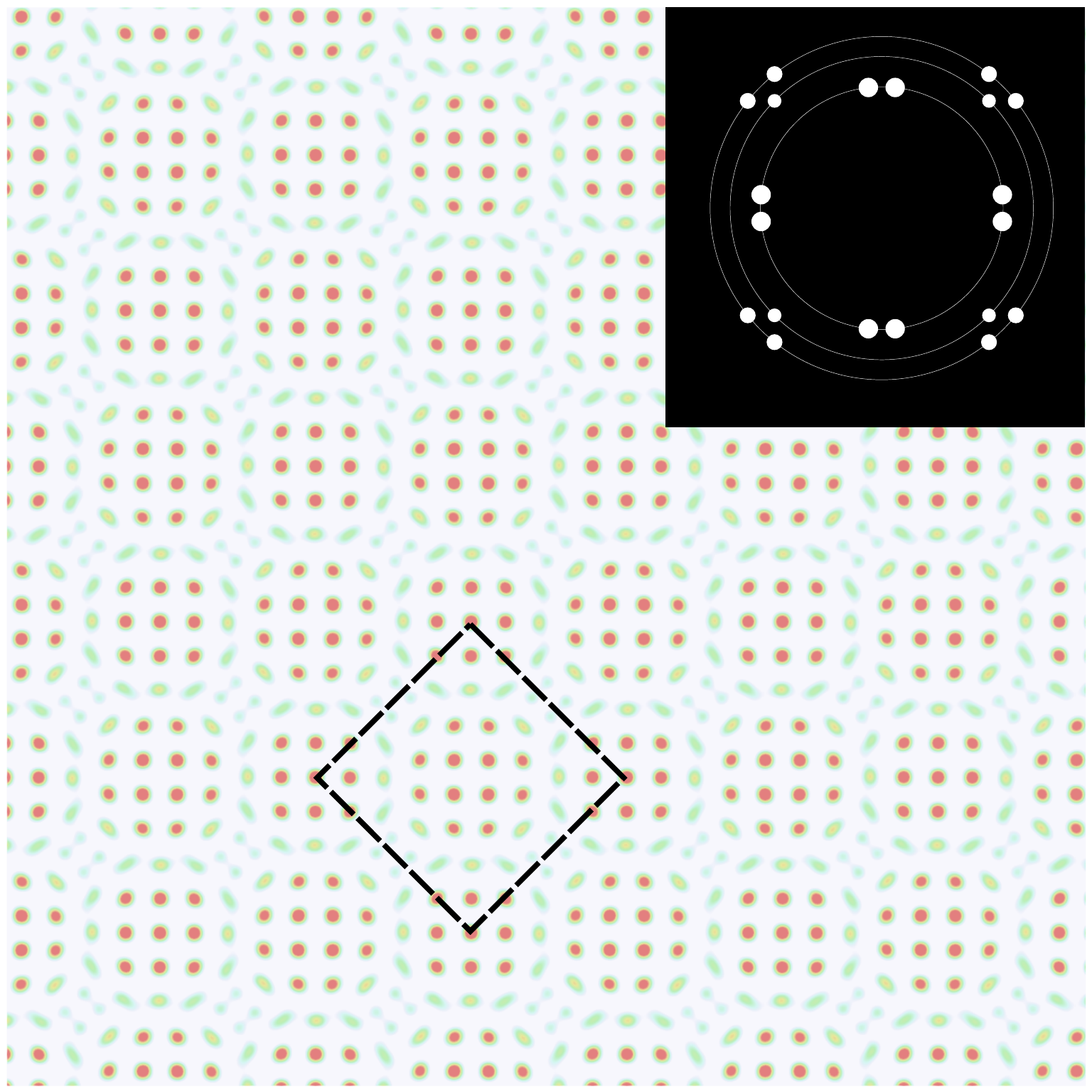}
	    \includegraphics[width=0.3\textwidth]{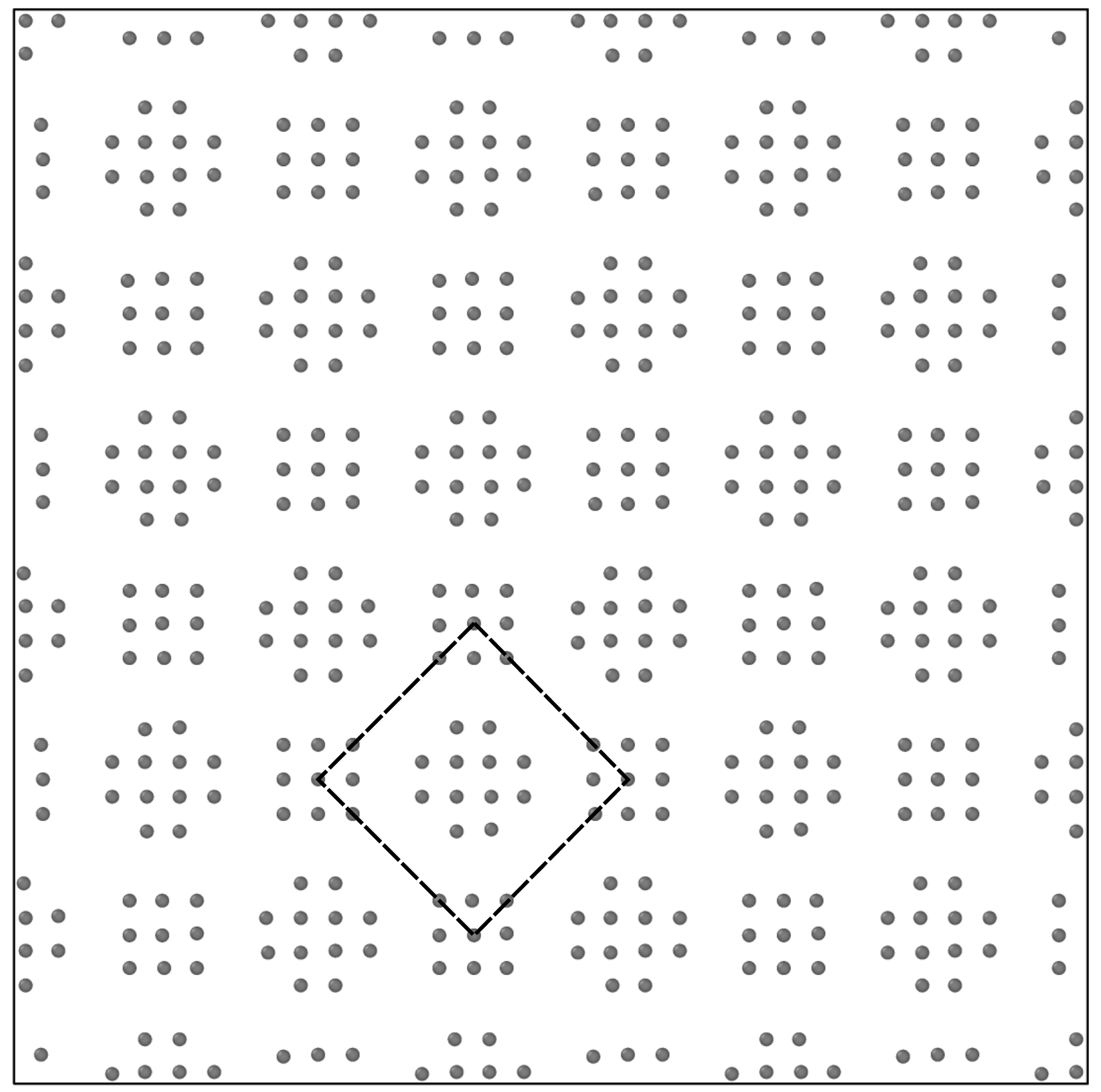}}
	\caption{Periodic $[100]$ twist GBs in BCC. (a) $\Sigma$5 GB with $\theta = 2\arctan(1/3) \approx 36.87^\circ$; (b) $\Sigma$41 GB with $\theta = 2\arctan(1/9) \approx 12.68^\circ$, accompanied by the atomic configuration.
	Dashed lines plotted on the diagram mark their periods.
	The distributions of the Fourier-Bohr spectra for the first three intensity levels are displayed in the upper right corner.
    }
	\label{fig:periodic_gb}
\end{figure*}

\section{Results}
\label{sec:results}
We present several interface examples to demonstrate the effectiveness of our model, including BCC $[100]$ twist and symmetric tilt GBs and BCC/FCC interfaces. To elucidate the structural features of the interfaces in our PFC results, we provide density maps of the probability scalar field $\varphi(\bm{r})$ evaluated at the cross-section $x = 0$. In these maps, darker regions rendered in red indicate higher values of $\varphi(\bm{r})$, while brighter regions rendered in light blue represent lower values, see \cref{fig:periodic_gb} for quantitative interpretation. The spacial distribution of $\varphi(\bm r)$ reflects statistically favorable atomic positions. Non-spherical regions observed in the density maps are atoms exhibiting increased spatial dispersion in their probability distributions, arising from local atomic mobility or structural flexibility near the interface.

To reconstruct the atomic configurations at the interface, we apply a threshold to $\varphi(\bm{r})$, yielding a level set as well as the high probability peaks where atoms are most likely located. The density $\rho$ of this thresholded level set is then computed to determine the associated tolerance parameter $\varepsilon$ in PCPS theory via \cref{formula:densrela}. A smaller value of $\varepsilon$ a higher threshold in the numerical results, thereby including fewer atoms in the predicted atomic structure and yielding a more precise structural prediction. In all visualizations presented, a 70\% threshold is employed.

\subsection{BCC $[100]$ twist GBs}
\label{sec:twist}
In this section, our model is employed to investigate the structural characteristics of BCC $[100]$ twist GB for varying twist angles $\theta$. Since the rotation is only applied on $y$-$z$ plane, the cut process in PCPS construction requires 
\begin{align}
    |x_1-x_2| \leq \varepsilon ,\ \frac{|x_1+x_2|}{2} \leq \varepsilon.
\end{align}
Since $x_1$ and $x_2$ are integers, these constraints imply $x_1=x_2=0$.
Consequently, the standard PCPS $\Lambda$ of BCC $[100]$ twist GB simplifies to a single CPS set projected from a 4 dimensional lattice to $\mathbb{R}^2$. 

Let $a$ denote the lattice constant of the BCC lattice in both bulk phases, and $\bm R_\theta$ be the canonical rotational matrix in the $y$-$z$ plane
\begin{align}
    \bm R_\theta  =\begin{pmatrix}
    \cos \theta & -\sin \theta \\ 
    \sin \theta & \cos \theta
    \end{pmatrix}.
\end{align}
The bulk lattice vectors of 2 grains are then represented as $\bm A = a\bm I_2,\bm B = a\bm R_\theta$. Accordingly, $\Lambda$ is constructed from 4 dimensional lattice
\begin{align}
\Gamma := \left\{ 
a\left(\begin{array}{l}
\bm I \quad \ -\bm R_\theta \\
\bm I/2 \quad \ \bm R_\theta/2
\end{array}\right)
\begin{pmatrix}
\bm{x} \\
\bm{y}
\end{pmatrix} 
\ \Big| \ \begin{pmatrix}
\bm{x} \\
\bm{y}
\end{pmatrix}   \in \mathbb{Z}^4 
\right\}`, 
\end{align}
where $\bm x= (y_1,z_1)^T,\bm y = (y_2,z_2)^T$, with the cut window specified as $W = \overline{B(\bm 0, \varepsilon)} \subset \mathbb{R}^2$. In numerical implementation, it is often preferable to assign symmetric rotations $\pm \theta/2$ to the two grains rather than applying a full rotation $\theta$ to one of them. In this convention, we have $\bm A=a\bm R_{-\theta/2},\bm B = a\bm R_{\theta/2}$. The Fourier-Bohr spectrum of resulting PCPS is then given by $\{\bm P \bm k \mid \bm k \in \mathbb{Z}^4\},\ \bm P = a^{-1}(\bm R_{\theta/2},\bm R_{-\theta/2})$. The parameters in the LB model (\ref{LB}) are chosen as $\alpha = 0.0,\ \gamma= 0.5$, within the regime where BCC phase is known to be stable \cite{mcclenagan2019landau}.

Based on classical CSL theory and direct visual observation, GB structures can be broadly classified into periodic and quasiperiodic cases. As we have discussed in \cref{sec:CPS}, when $\tan ({\theta}/{2})$ is a rational number, twist GBs are periodic. Otherwise when $\tan ({\theta}/{2})$ is irrational, our PCPS model reveals that the resulting GB structure is quasiperiodic. 

\subsubsection{Periodic twist GBs}
We demonstrate periodic GBs corresponding to both low- and high-$\Sigma$ values, see \cref{fig:periodic_gb}. Prior studies on BCC periodic GBs can be found in \cite{scheiber2016ab, fu2020optical, feng2015energy, frolov2018grain}. Specifically, we choose $\theta = 2\arctan(1/3)$ ($\Sigma$5) as a representative low-$\Sigma$ case and $\theta = 2\arctan(1/9)$ ($\Sigma$41) as a representative high-$\Sigma$ case. The Fourier-Bohr spectra of both cases has the form $\{\bm{P}\bm{k}\mid \bm{k}\in \mathbb{Z}^2\}$, where projection matrix $\bm{P}\in \mathbb{R}^{2 \times 2}$. As discussed in \cref{sec:PFC}, these spectra are the reciprocal lattices of the underlying physical periodicity.

For these specific twist angles, the resulting interfaces exhibit well-defined periodic structures. Distinct CSL supercells are clearly discernible and are highlighted. Atoms occupying coincidence sites are located at the vertices of the black rectangles, which consistently coincide with the high-density regions in $\varphi(\bm{r})$. Furthermore, atoms at non-coincidence but PCPS sites also produce locally high-density regions that are visually indistinguishable from those at coincidence sites. In the high-$\Sigma$ case of $\theta = 2 \arctan(1/9)$, periodic dislocation network could be observed, and similar structures could also emerge in quasiperiodic case, see \cref{sec:lagb}. These phenomena illustrate the periodic arrangement of atomic positions and the high crystallographic order of periodic interfaces.

\subsubsection{Quasiperiodic twist GBs}
Based on the GB pattern profiles and prior studies across various crystallographic systems \cite{feng2015energy, fu2020optical, dai2014atomistic, yang2010atomic}, twist GBs can be broadly categorized into distinct types depending on the twist angle $\theta$. Although different classification schemes, such as those based on GB energy, dislocation density, or structural units, may yield varying categorizations, the dependence of interface properties on twist angle is generally continuous, with the exception of sharp energy reductions at certain low-$\Sigma$ commensurate twist angles \cite{scheiber2016ab, feng2015energy, zou2025quasiperiodic}.

\begin{figure}[!hpbt]
	\centering
	\includegraphics[width=0.4\textwidth]{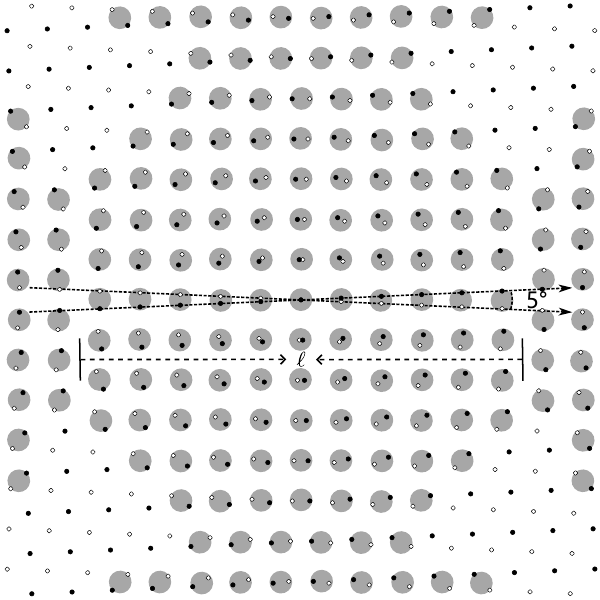}
	\caption{The PCPS diagram for $\theta = 5^\circ$. Large grey circles of diameter $\varepsilon$ are generated by black and white circles with distances less than $\varepsilon$. Blockwise atoms having ordered orientations separated by dislocation networks are observed with variable side length with approximation $\ell \approx a/\tan\theta \approx 11.43a$.
	}
	\label{fig:arctanrela}
\end{figure}

\begin{figure*}[!hpbt]
	\centering
	\subfigure[$\theta = 2^\circ$]{
		\includegraphics[width=0.32\textwidth]{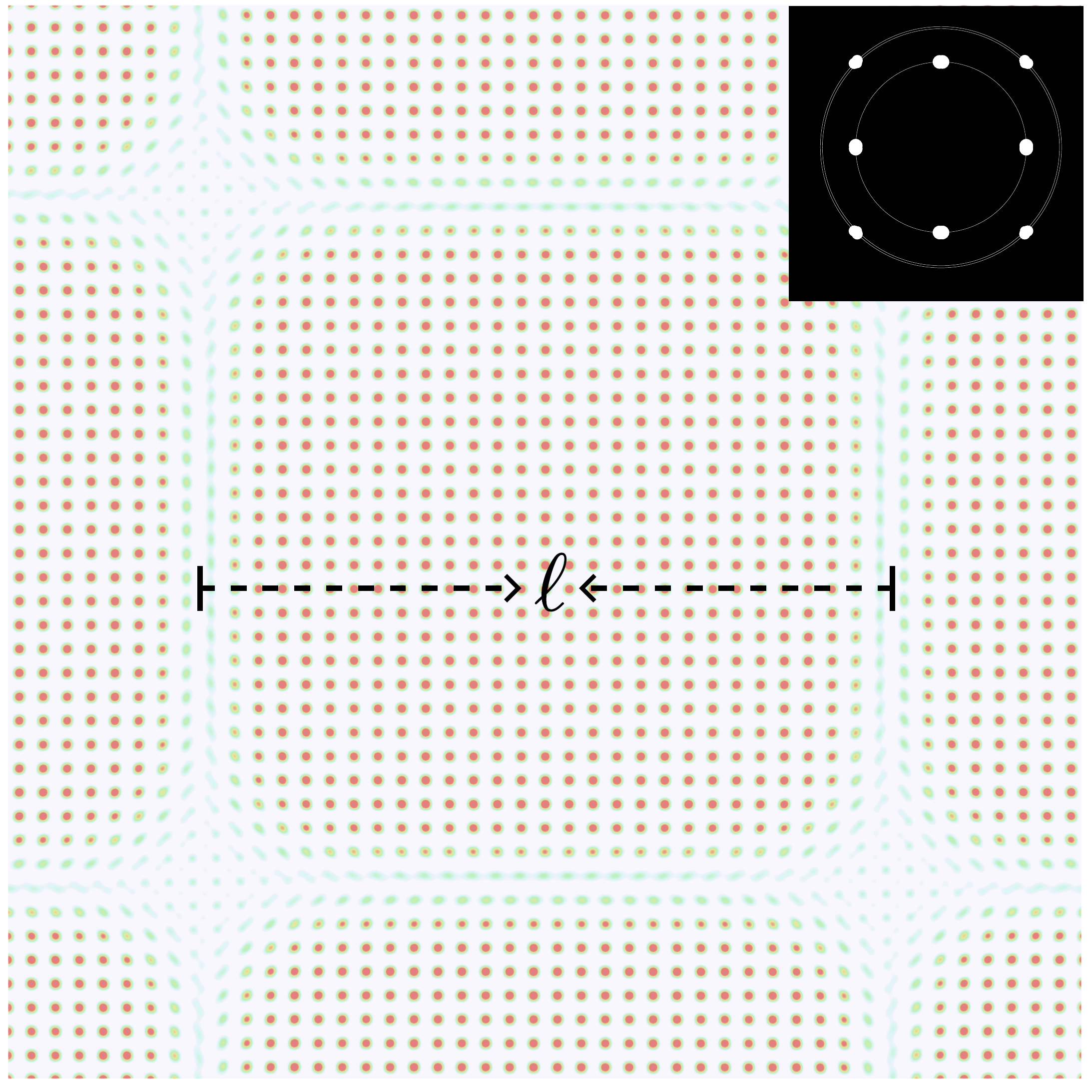}}
	\subfigure[$\theta = 5^\circ$]{
		\includegraphics[width=0.32\textwidth]{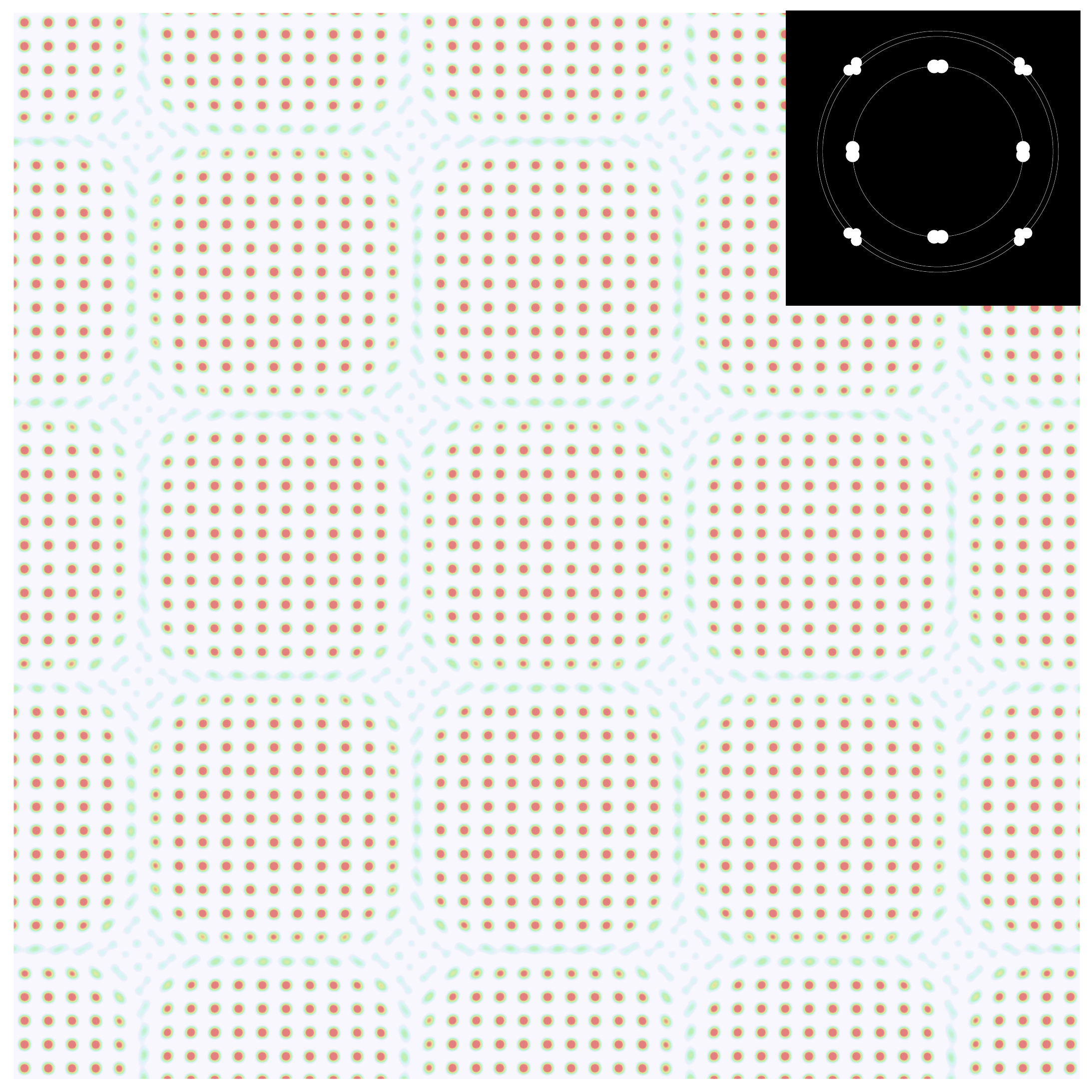}
		\label{fig:low_angle_qgb_5}}
	\subfigure[$\theta = 9^\circ$]{
		\includegraphics[width=0.32\textwidth]{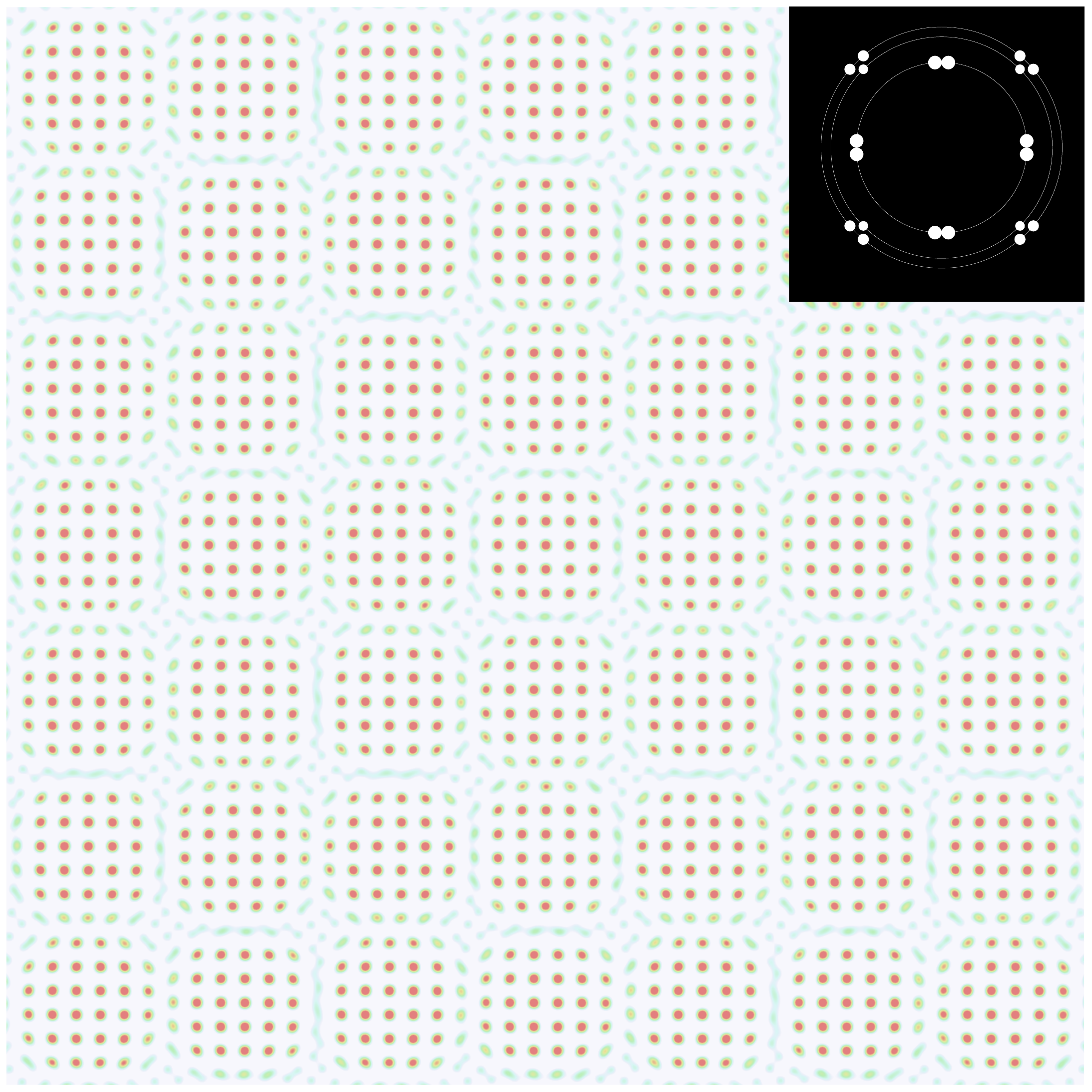}}
	\subfigure[$\theta = 12^\circ$]{
		\includegraphics[width=0.32\textwidth]{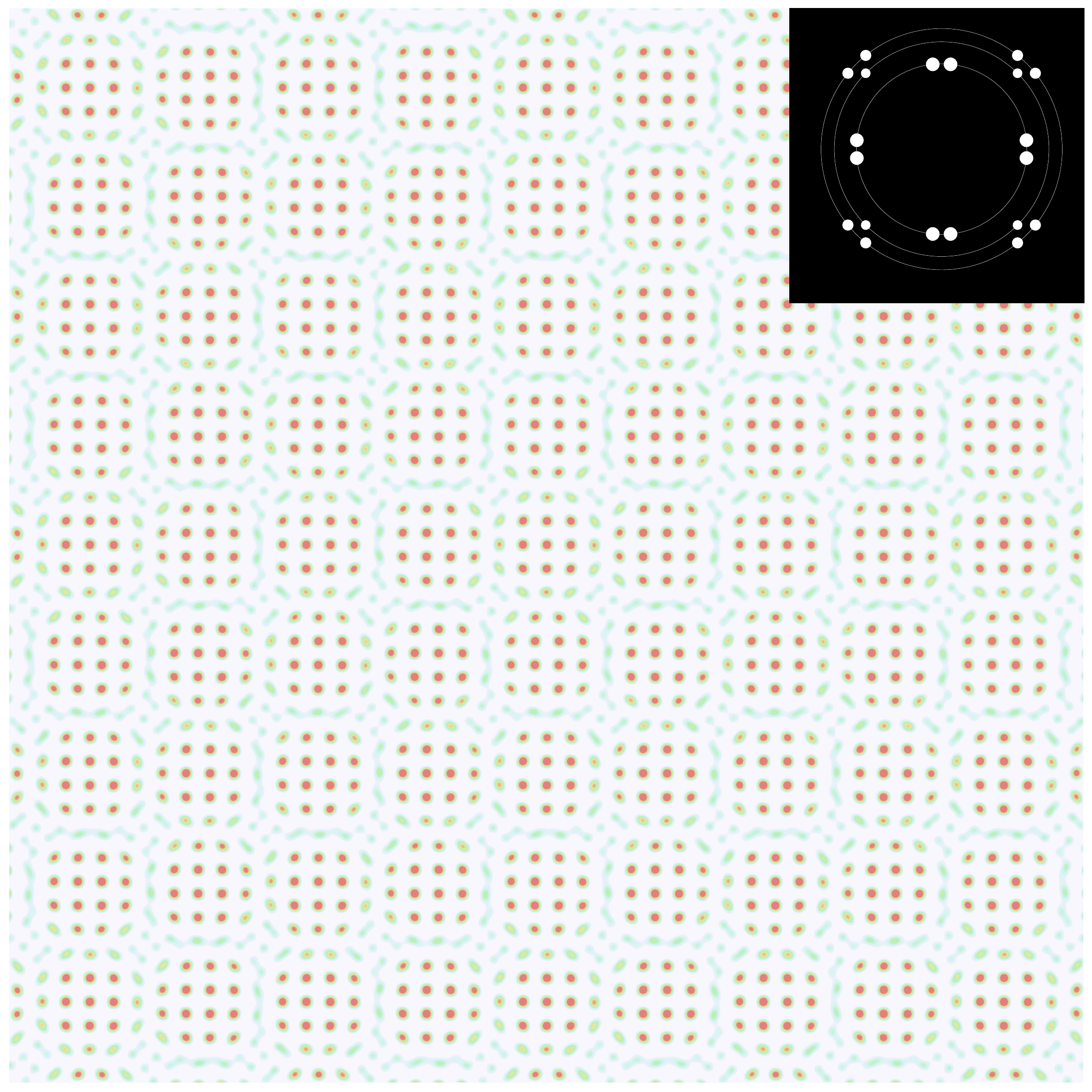}}
	\subfigure[$\theta = 15^\circ$. Thresholded level set and atomic configuration.]{
		\includegraphics[width=0.32\textwidth]{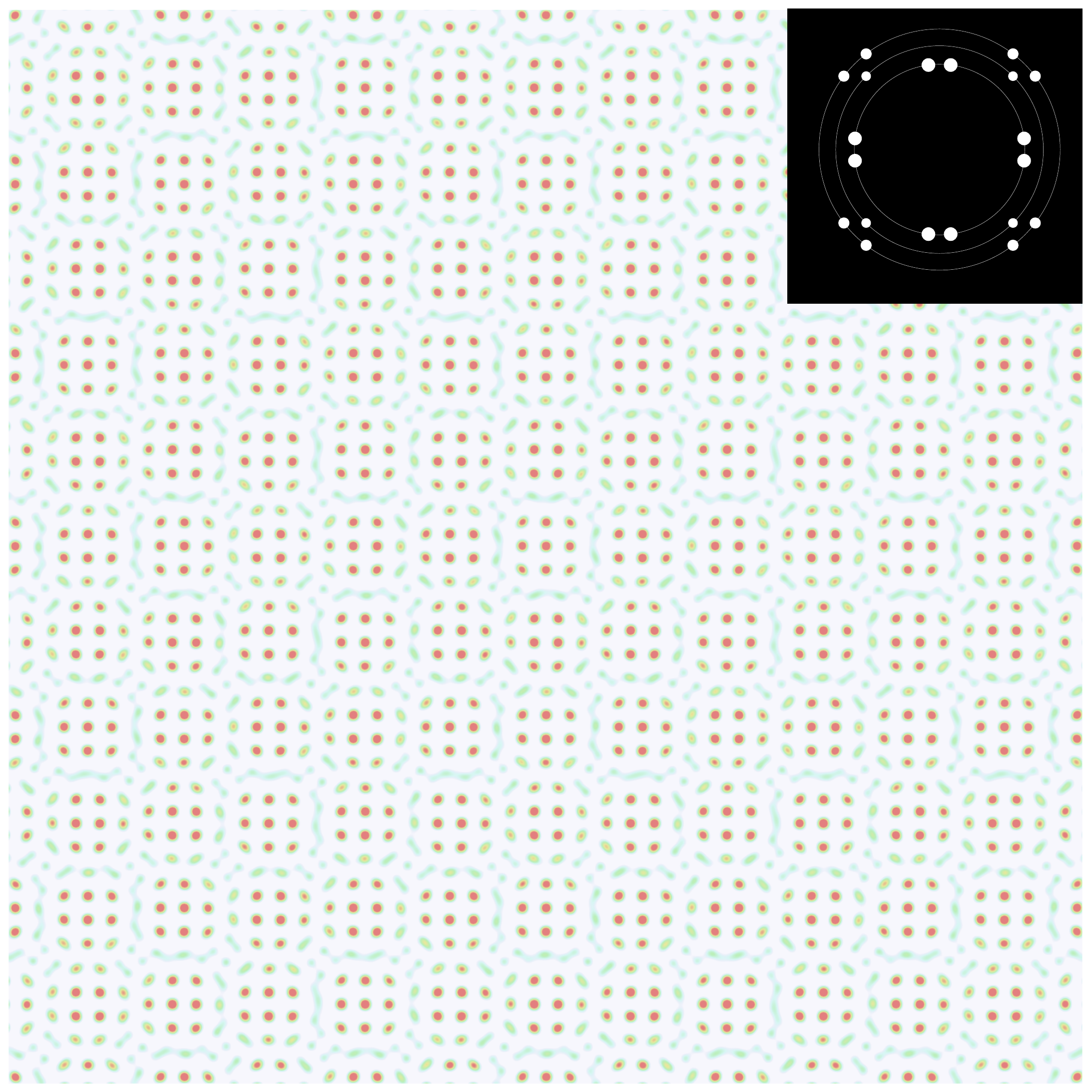}
		\includegraphics[width=0.32\textwidth]{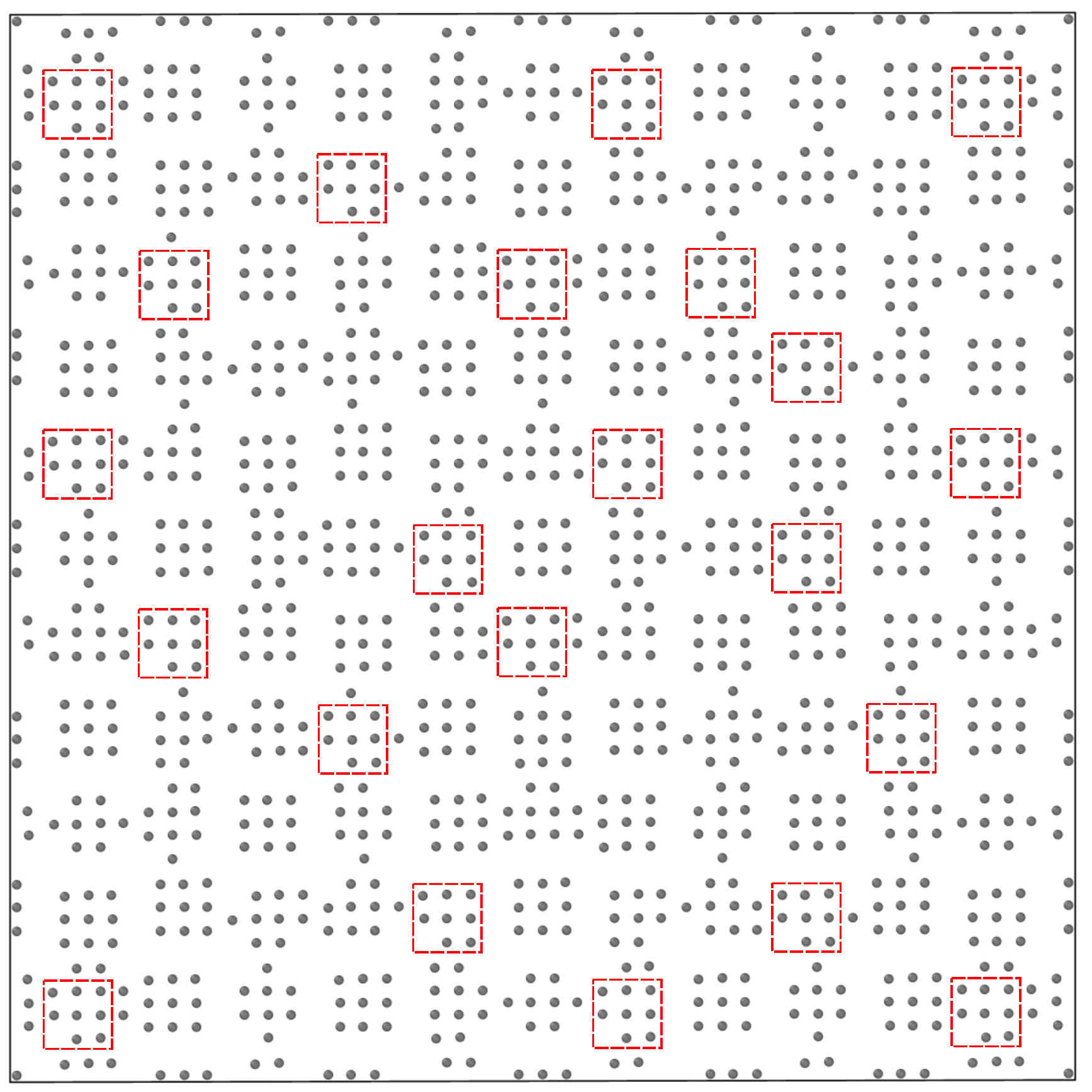}
		\label{fig:low_angle_qgb_15}}
	\caption{Low-angle quasiperiodic $[100]$ twist GBs in BCC: (a) $\theta = 2^\circ$; (b) $\theta = 5^\circ$; (c) $\theta = 9^\circ$; (d) $\theta = 12^\circ$; (e) $\theta = 15^\circ$. Dislocation networks separating high-density blocks are clearly visible. The orientations of high-density blocks and GB pattern patches repeat in a quasiperiodic manner. The red dashed-line rectangles in (e) highlight identical local patterns, confirming the presence of repetitivity. Other quasiperiodic features such as FLC are also visually evident. The distributions of the Fourier-Bohr spectra for the first three intensity levels are displayed in the upper right corner.}
	\label{fig:low_angle_qgb}
\end{figure*}

In this work, we adopt a classification scheme based primarily on the geometric patterns of the GB, that is, the spatial distribution of atoms across the interface. Through our observations of the distinct patterns shown in \cref{fig:periodic_gb} for periodic GBs, the resulting structures can be emprically categorized into two principal regimes: low-angle and high-angle GBs, in line with previous works \cite{feng2015energy}. While these regimes are typically distinguished by different structural features, it is notable that apart from a few special commensurate twist angles, the GB structures in both regimes are quasiperiodic.

\paragraph{Low-angle GBs}
\label{sec:lagb}
When $\theta \lesssim 15^\circ$, most atoms within the bulk phases remain locally fitted, while a subset of atoms  exhibit misfits, forming blockwise distributed $\varepsilon$-proximal pairs as illustrated in \cref{fig:arctanrela}. This process leads to the development of a well-defined dislocation network. The PCPS model also provides a estimation for the variable side length of the dislocation network. By evaluating the spatial distance necessary for adjacent grains to accommodate the misorientation, the side length $\ell$ varies around
\begin{align}
    \ell \approx a / \tan\theta.
\end{align}
This relation is supported by the data presented in \cref{fig:arctanrela} and \cref{fig:low_angle_qgb_5}. In \cref{fig:arctanrela}, $\varepsilon$ is chosen as $\varepsilon \approx 0.552a$ to match the density of the thresholded level set. Theoretical predictions exhibit excellent agreement with the numerical results.

We present several low-angle GBs for quasiperiodic twist angles in \cref{fig:low_angle_qgb}, corresponding to $\theta = 2^\circ,\ 5^\circ,\ 9^\circ,\ 12^\circ$ and $15^\circ$. The Fourier-Bohr spectra of these quasiperiodic cases has the form $\{\bm{P}\bm{k}\mid \bm{k}\in \mathbb{Z}^{4}\}$ where $\bm{P}\in \mathbb{R}^{2\times 4}$. The first and second level spectral components are all correspond to \cref{tab:twist_spectra}. 
\begin{table}[!htbp]
    \centering
    \caption{Spectral components of the first and second levels of intensity in BCC $[100]$ quasiperiodic twist GBs.}
    \label{tab:twist_spectra}
    \begin{tabular}{cc}
        \toprule
        ~~Intensity~~ & Spectra $\bm P \bm k$ \\
        \midrule
        \multirow{2}{*}{1st level} & $\bm{P}(\pm 1,0,0,0)^T,\ \bm{P}(0,\pm 1,0,0)^T$ \\
        & $\bm{P}(0,0,\pm 1,0)^T,\ \bm{P}(0,0,0,\pm 1)^T$ \\
        \\[-0.5em]
        2nd level & $\bm{P}(\pm 1,\pm 1,0,0)^T,\ \bm{P}(0,0,\pm 1,\pm 1)^T$ \\
        \bottomrule
    \end{tabular}
\end{table}
Quasiperiodicity is confirmed through $\mathbb{Q}$-independence of these spectral components. Other representative quasiperiodic features, including FLC and repetitivity, are also visually evident in thresholded level set \cref{fig:low_angle_qgb_15}. Similar repetitive characteristics can be seen in \cite{demkowicz2008interfaces}. Both numerical observations and theoretical predictions indicate that the characteristic size of the dislocation network decreases with increasing twist angle $\theta$, in accordance with the proportional behavior of $1/\tan \theta$. 
This trend is consistent with classical theories such as the O-lattice model \cite{bollmann1982crystal}. Within each block, atomic positions represented by the local maxima of the density field $\varphi(\bm{r})$ exhibit nearly periodic order disrupted only at dislocation lines. The local orientation of atomic patterns within each block is also predictable within the framework of our model. 
The inter-block regions, characterized by significantly lower density, serve to accommodate the relatively uniform part of $\varphi(\bm{r})$. Density distributions in these regions display lower local peaks and broader spatial spread, thus exhibiting greater atomic mobility compared to those correspond with proximal pairs.

Previous studies in low-angle twist GBs mainly focused on periodic cases, observing periodic dislocation networks \cite{fu2020optical, feng2015energy}.
In this work, we are able to obtain the global visual and spectral characteristics of the system, allowing a more accurate assessment of the interface patterns.
Combined with the PCPS theoretical framework, our results indicate that low-angle twist GBs are inherently quasiperiodic. However, the quasiperiodicity in low-angle GBs is relatively subtle and typically becomes apparent only in large-scale visualizations. As the twist angle increases, atomic mismatch grows, thereby enhancing the manifestation of quasiperiodicity across the interface.

\paragraph{High-angle GBs}

\begin{figure*}[!hpbt]
	\centering
	\subfigure[$\theta = 18^\circ$]{
		\includegraphics[width=0.32\textwidth]{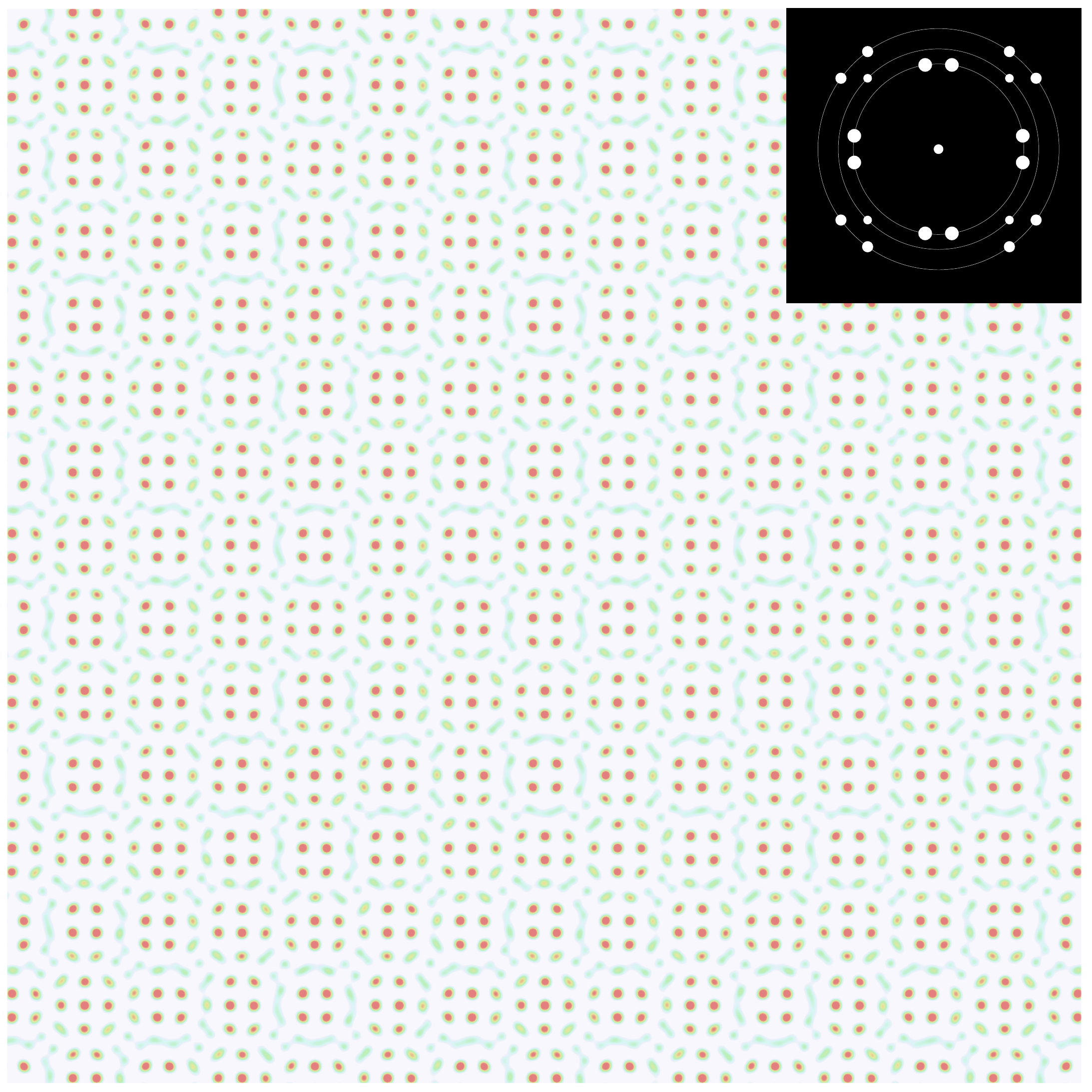}}
	\subfigure[$\theta = 24^\circ$]{
		\includegraphics[width=0.32\textwidth]{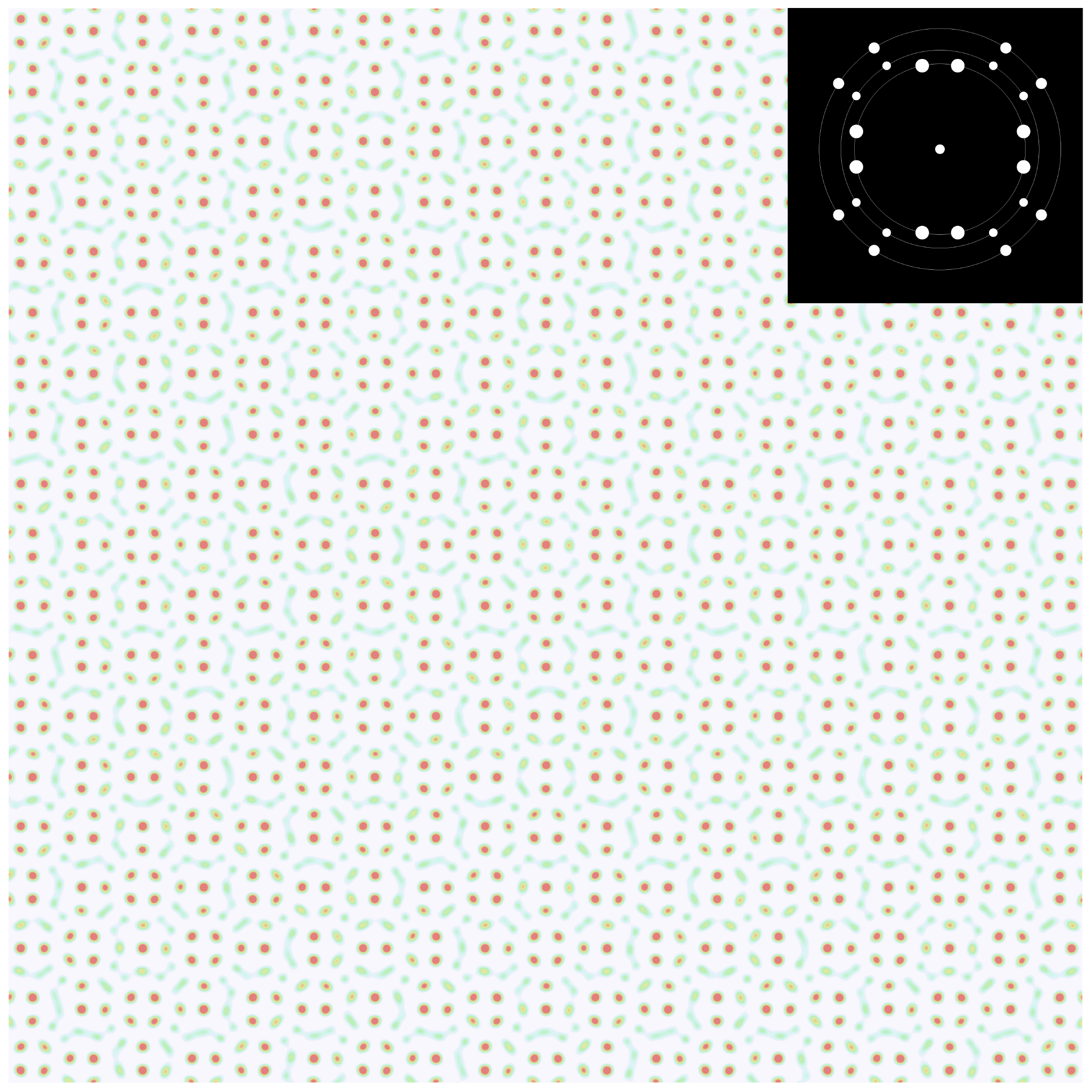}}
    \subfigure[$\theta = 36^\circ$]{
		\includegraphics[width=0.32\textwidth]{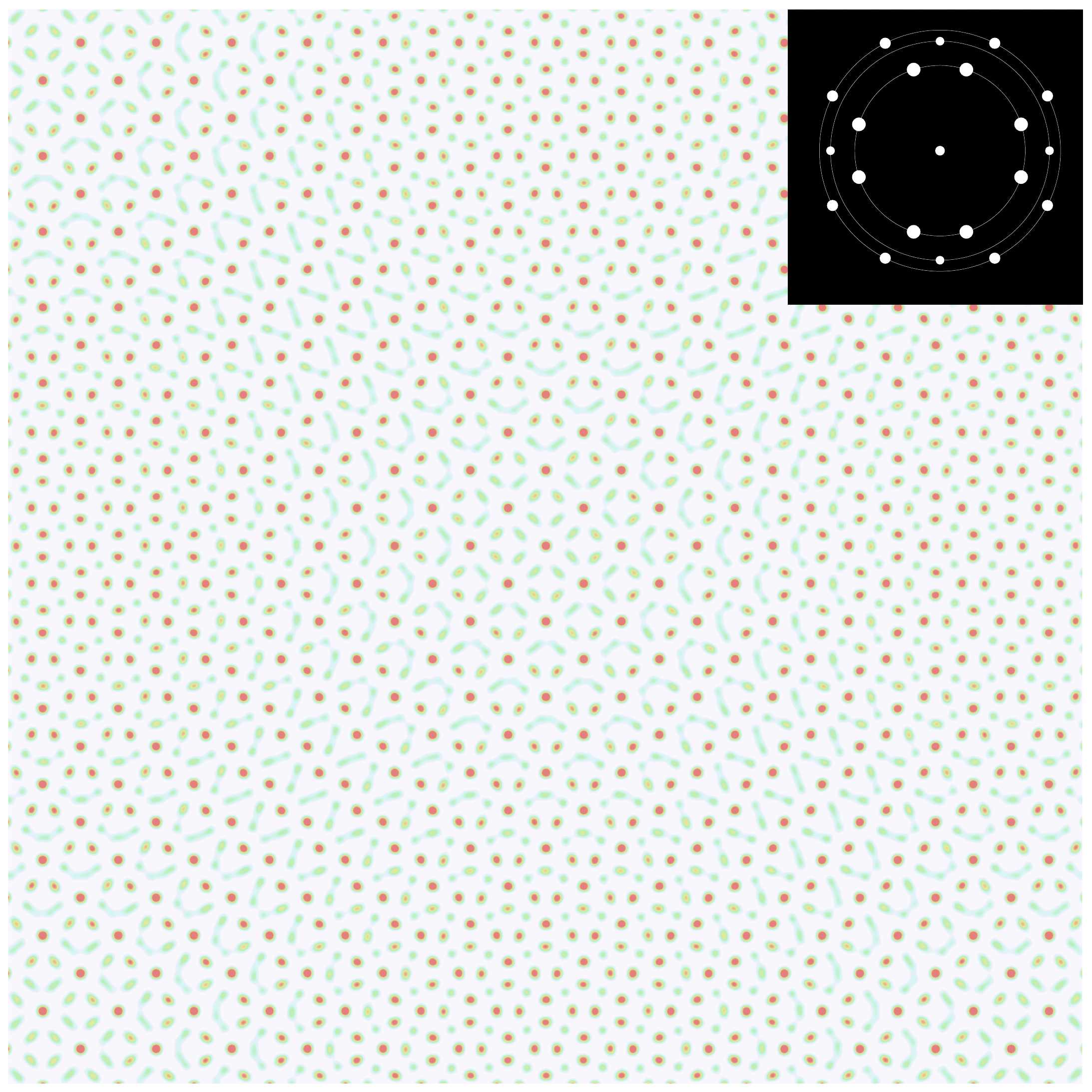}}
	\caption{High-angle quasiperiodic $[100]$ twist GBs in BCC: (a) $\theta = 18^\circ$; (b) $\theta = 24^\circ$; (c) $\theta = 36^\circ$.
	 The distributions of the Fourier-Bohr spectra for the first three intensity levels are displayed in the upper right corner.}
	\label{fig:high_angle_qgb_1}
\end{figure*}

\begin{figure*}[!hpbt]
	\centering
	\subfigure[$\theta = 30^\circ$. Thresholded level set and atomic distribution.]{
		\includegraphics[width=0.4\textwidth]{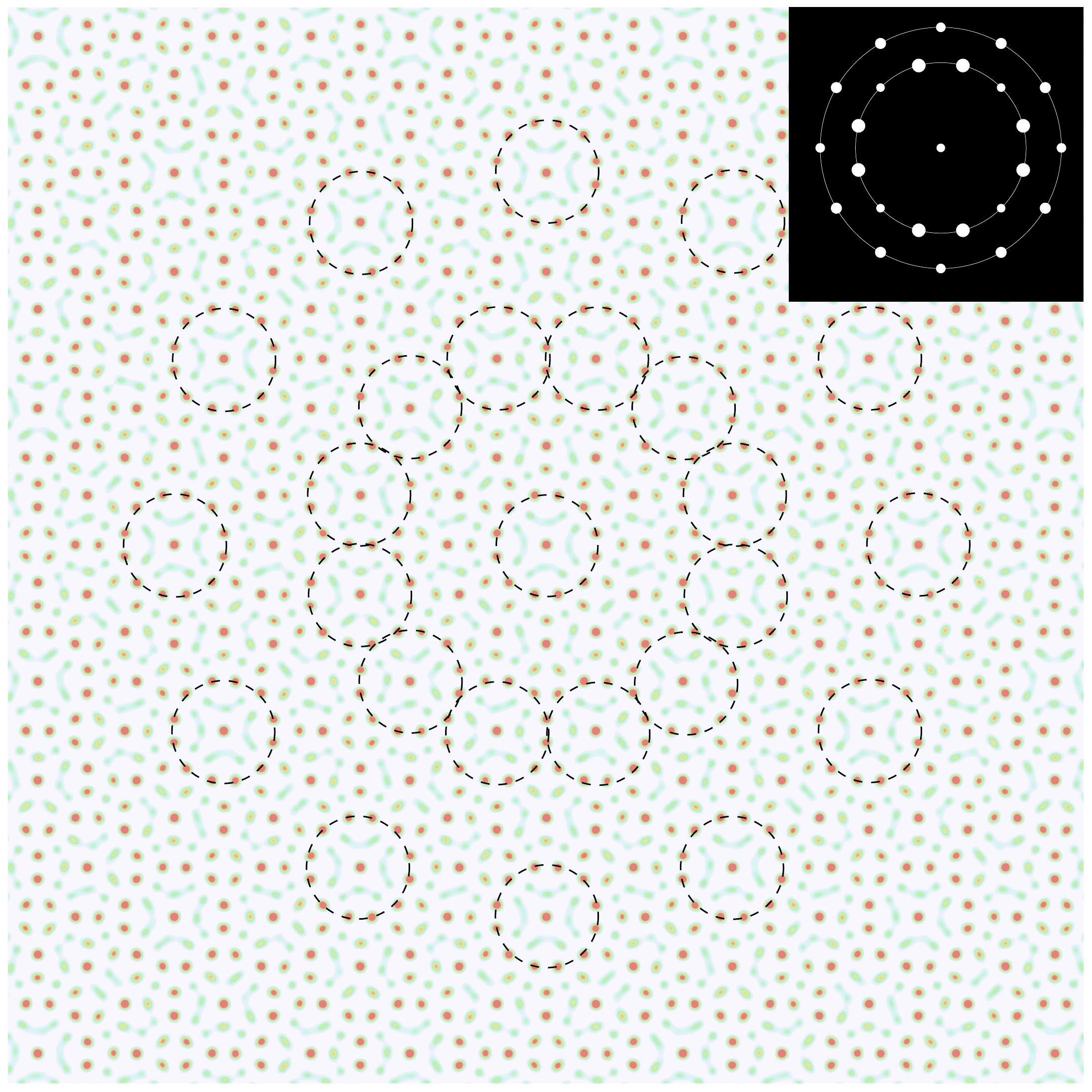}
		\includegraphics[width=0.4\textwidth]{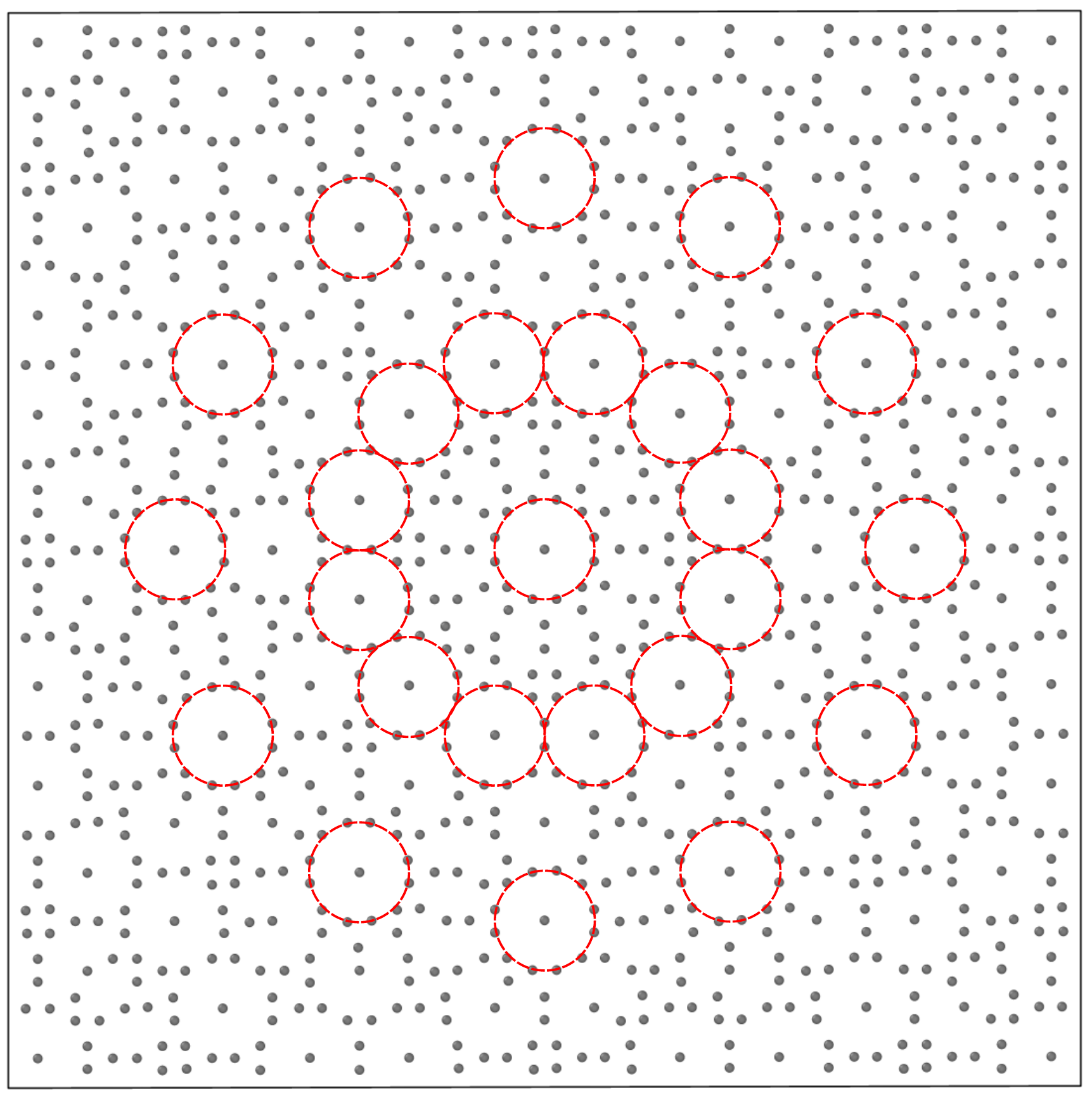}}
		\\
	\subfigure[$\theta = 45^\circ$. Thresholded level set and atomic distribution.]{
		\includegraphics[width=0.4\textwidth]{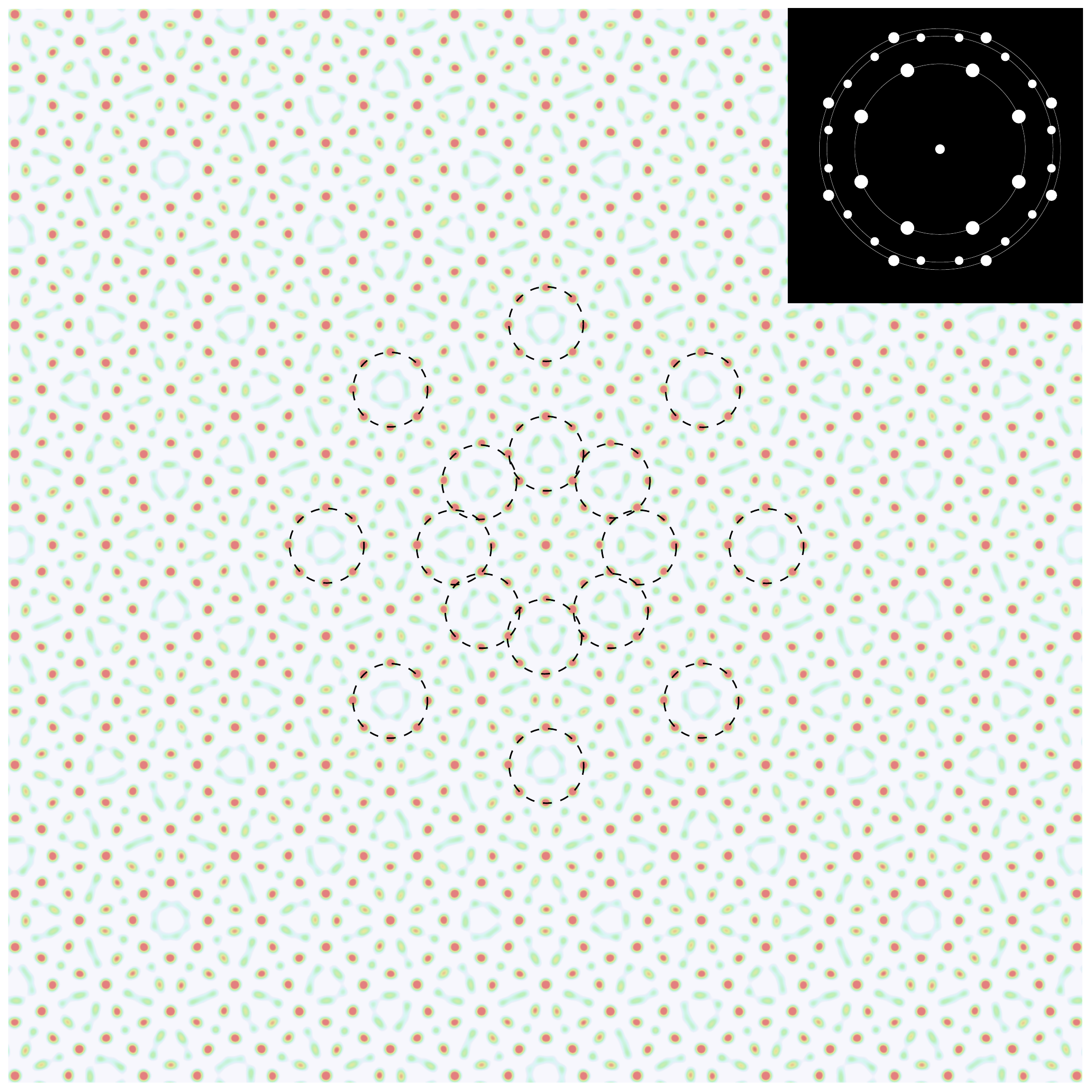}
		\includegraphics[width=0.4\textwidth]{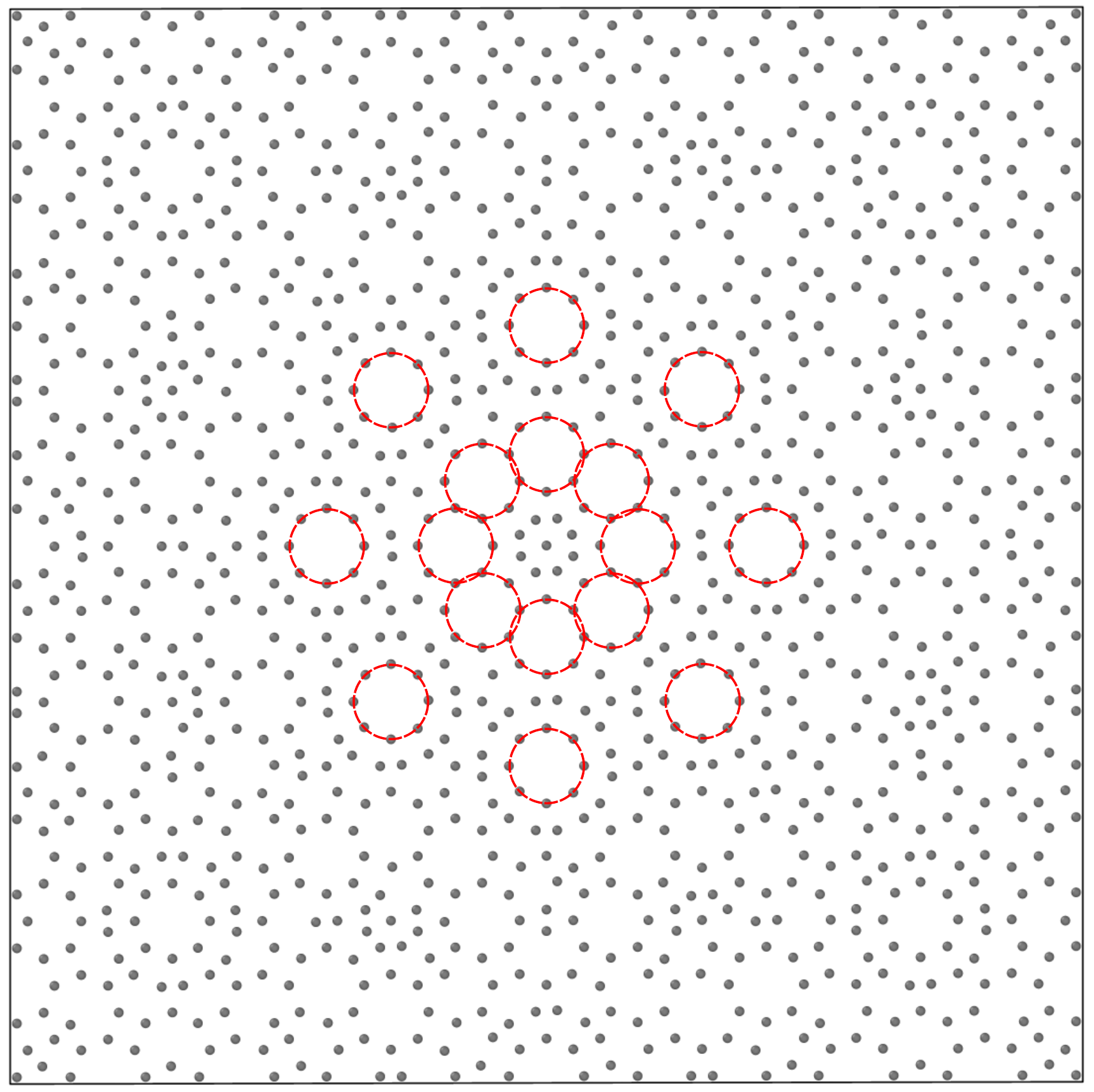}}
	\caption{High-angle quasiperiodic $[100]$ twist GBs in BCC: (a) $\theta = 30^\circ$ and (b) $\theta = 45^\circ$. Identifiable 12- and 8-fold symmetric patterns are demonstratively enclosed by dashed-line dodecagons and octagons respectively. Other quasiperiodic features such as FLC and repetitivity can also be confirmed, consistent with the observations in \cref{fig:low_angle_qgb_15}.
	}
	\label{fig:high_angle_qgb_2}
\end{figure*}

Dislocation networks observed in low-angle twist GBs become disordered and gradually disappear when $\theta \gtrsim 15^\circ$, as illustrated in \cref{fig:high_angle_qgb_1} and \cref{fig:high_angle_qgb_2}. This breakdown can be attributed to the widening spatial separation between interacting atomic layers. In this case, quasiperiodicity becomes significantly more apparent, particularly when the interface pattern is examined over a large spatial domain.

We select several high-angle twist GBs to present our numerical results, with a particular focus on angles that are rational multiples of $\pi$; namely, $\theta = 18^\circ,\ 24^\circ,\ 30^\circ,\ 36^\circ$, and $45^\circ$ in our study. No regular dislocation network is observed under large-scale inspection in these cases. The breakdown of periodicity is evident both in spacial atomic configurations and in the corresponding Fourier-Bohr spectra distribution, see \cref{fig:high_angle_qgb_1}. Simultaneously, quasiperiodic characteristics such as FLC and repetitivity remain prominent across all configurations.

Another noteworthy phenomenon appears at the special twist angles \( \theta = 30^\circ \) and \( \theta = 45^\circ\), see \cref{fig:high_angle_qgb_2}. Despite the presence of general quasiperiodic features, one can visually identify the 12- and 8-fold symmetries in the numerical results for \( 30^\circ\) and \( 45^\circ\) cases. These rotational symmetries are hallmark characteristics of quasicrystals and are consistent with phenomena in other quasicrystalline materials \cite{zou2025quasiperiodic, ShechtmanMet, yin2021transition}. A straightforward mathematical observation regarding the occrurence of quasicrystals is that both $30^\circ=2\pi/12$ and $45^\circ = 2\pi/8$ are of the form $2\pi/n$, where $n\in \mathbb{N}$. However, having a twist angle of the form $2\pi/n$ is not a sufficient condition for the emergence of non-crystallographic rotational symmetries. For instance, our numerical results for $\theta = 18^\circ$ (equivalent to $72^\circ= 2\pi/5$), $\theta=36^\circ = 2\pi/10$ do not display such rotational symmetry. We will explore the origins and implications of these non-crystallographic rotational symmetries in more detail in the next section.

\subsubsection{Non-crystallographic rotational symmetries}
\label{sec:qc}
In our numerical results, 12- and 8-fold rotational symmetries are empirically observed when the twist angle $\theta = 30^\circ$ and $\theta = 45^\circ$ respectively. In contrast, other non-crystalline rotational symmetries, such as 5- and 10-fold, are absent. The PCPS theoretical framework developed accurately predicts and explains these phenomena through the following claim: for arbitrary $\varepsilon$, the PCPS $\Lambda$ supports only vague 12- and 8-fold symmetries up to $\varepsilon/2$-perturbation, corresponding to $\theta = 30^\circ$ and $\theta = 45^\circ$ respectively.

The phrases ``vague 12-fold/8-fold symmetries" and ``up to $\varepsilon/2$-perturbation" should be understood as a sort of visual indistinguishability, described as follows: there exists a point set $\Lambda'$, which is itself a PCPS and thus $\varepsilon/2$-close to $\Lambda$ (in the sense of \textit{vague topology}), that exhibits rigorous 12- or 8-fold rotational symmetry. The proof of this result hinges on the symmetry of the set $\pi_{\perp}^{-1}(W) \cap \Gamma$ in the internal space. Since the proof for $\theta = 45^\circ$ follows analogously to the $\theta = 30^\circ$ case, we present only the latter. For simplicity, the BCC lattice constant is set to $a=1$.

As discussed in \cref{sec:CPS} and \cref{formula:CPS}, the CPS for generating standard PCPS $\Lambda$ involves a 2 dimensional window \( W = \overline{B(\boldsymbol{0}, \varepsilon)} \) and a 4 dimensional lattice defined as
\begin{align}
\footnotesize
    \Gamma := \left\{ 
    \left( \begin{matrix}
        1 & 0 & -\sqrt{3}/{2} & {1}/{2}  \\
        0 & 1 & -{1}/{2} & -{\sqrt{3}}/{2} \\
       {1}/{2} & 0 & {\sqrt{3}}/{4} & -{1}/{4} \\ 
       0 & {1}/{2} &  {1}/{4} & {\sqrt{3}}/{4}
    \end{matrix} \right)
    \left( \begin{matrix}
        y_1 \\
        z_1 \\
        y_2 \\
        z_2
    \end{matrix} \right) 
    \ \Bigg| \ \left( \begin{matrix}
        y_1 \\
        z_1 \\
        y_2 \\
        z_2
    \end{matrix} \right) \in \mathbb{Z}^4 
    \right\}.
\end{align}
The projection of $\Gamma$ into internal space, $\pi_{\perp}(\Gamma)$, is given by
\begin{align}
\footnotesize
\begin{split}
 \pi_{\perp}(\Gamma) &= \left\{ 
 \left( \begin{matrix}
        1 & 0 & -\sqrt{3}/{2} & {1}/{2}  \\
        0 & 1 & -{1}/{2} & -{\sqrt{3}}/{2} \\
 \end{matrix} \right)
 \left( \begin{matrix}
    y_1 \\
    z_1 \\
    y_2 \\
    z_2
 \end{matrix} \right) 
 \ \Bigg| \ \left( \begin{matrix}
        y_1 \\
        z_1 \\
        y_2 \\
        z_2
    \end{matrix} \right) \in \mathbb{Z}^4 
 \right\} \\
 &= \mathbb{Z}[\xi_{12}],
 \end{split}
\end{align}
where $\mathbb{Z}[\xi_{12}]$ denotes the ring of integers in the cyclotomic field $\mathbb{Q}(\xi_{12})$ \cite{washington2012introduction}, which is known to possess 12-fold rotational symmetry. Given that the window $W$ is a ball centered at the origin, the set of internal space points corresponding to $\Lambda$, namely \( W \cap \mathbb{Z}[\xi_{12}] \), also inherits this 12-fold symmetry.

 Now, consider any point $(y_1, z_1, y_2, z_2)$ such that
\begin{align}
    \left( \begin{matrix}
       1 & 0 & -\sqrt{3}/{2} & {1}/{2}  \\
        0 & 1 & -{1}/{2} & -{\sqrt{3}}/{2}
    \end{matrix} \right)
    \left( \begin{matrix}
        y_1 \\
        z_1 \\
        y_2 \\
        z_2
    \end{matrix} \right) 
    = \bm{v} \in \overline{B(\bm{0}, \varepsilon)},
\end{align}
then its rotation by ${\pi}/{6}$, denoted $\bm{R}_{{\pi}/{6}} \bm{v}$, also lies in $\overline{B(\bm{0}, \varepsilon)}$. One can show that this rotated vector admits the following representation
\begin{align}
    \left( \begin{matrix}
        1 & 0 & -\sqrt{3}/{2} & {1}/{2}  \\
        0 & 1 & -{1}/{2} & -{\sqrt{3}}/{2}
    \end{matrix} \right)
    \left( \begin{matrix}
        - y_2 \\
        - z_2 \\
        -z_2-y_1 \\
        y_2-z_1
    \end{matrix} \right) 
    = \bm{R}_{{\pi}/{6}} \bm{v} \in \overline{B(\bm{0}, \varepsilon)}.
\end{align}

Since both $\bm{v}$ and $\bm{R}_{{\pi}/{6}} \bm{v}$ are contained in $\overline{B(\bm{0}, \varepsilon)}$, there must exist points $\bm{p},\bm{q}\in \Lambda$ corresponding respectively to $\bm{v}$ and $\bm{R}_{{\pi}/{6}} \bm{v}$, having the form
\begin{align}
\bm{p} &= \frac{1}{2}\left( \begin{matrix}
       1 & 0 & \sqrt{3}/{2} & -{1}/{2}  \\
        0 & 1 & {1}/{2} & {\sqrt{3}}/{2}
    \end{matrix} \right)
    \left( \begin{matrix}
        y_1 \\
        z_1 \\
        y_2 \\
        z_2
    \end{matrix} \right),\\ 
\bm{q} &= \frac{1}{2}\left( \begin{matrix}
        1 & 0 & \sqrt{3}/{2} & -{1}/{2}  \\
        0 & 1 & {1}/{2} & {\sqrt{3}}/{2}
    \end{matrix} \right)
    \left( \begin{matrix}
        - y_2 \\
        - z_2 \\
        -z_2-y_1 \\
        y_2-z_1
    \end{matrix} \right).
\end{align}
The angle between $\bm{p},\bm{q}\in \Lambda$ turns out to be approximately \( {5\pi}/{6} \). This can be seen by noting that 
\begin{align}
\begin{split}
    \bm{R}_{{5\pi}/{6}}\bm{p} &= \frac{\bm{R}_{{5\pi}/{6}}}{2}\left( \begin{matrix}
       1 & 0 & \sqrt{3}/{2} & -{1}/{2}  \\
        0 & 1 & {1}/{2} & {\sqrt{3}}/{2}
    \end{matrix} \right)
    \left( \begin{matrix}
        y_1 \\
        z_1 \\
        y_2 \\
        z_2
    \end{matrix} \right)\\ 
    &=\frac{1}{2}\left( \begin{matrix}
       1 & 0 & \sqrt{3}/{2} & -{1}/{2}  \\
        0 & 1 & {1}/{2} & {\sqrt{3}}/{2}
    \end{matrix} \right)
    \left( \begin{matrix}
        -y_2-z_1 \\
        y_1-z_2 \\
        -y_1 \\
        -z_1
    \end{matrix} \right).
\end{split}
\end{align}
Then the distance between $\bm{R}_{{5\pi}/{6}}\bm{p}$ and $\bm{q}$ satisfies 
\begin{align}
\begin{split}
    \left|\bm{R}_{{5\pi}/{6}}\bm{p} - \bm{q}\right| &=\left| \frac{1}{2}\left( \begin{matrix}
       1 & 0 & \sqrt{3}/{2} & -{1}/{2}  \\
        0 & 1 & {1}/{2} & {\sqrt{3}}/{2}
    \end{matrix} \right) \left( \begin{matrix}
        -z_1 \\
        y_1 \\
        z_2 \\
        -y_2
    \end{matrix} \right)\right|\\ 
    &=\left| \frac{\bm{R}_{{\pi}/{2}}}{2}\left( \begin{matrix}
       1 & 0 & -\sqrt{3}/{2} & {1}/{2}  \\
        0 & 1 & -{1}/{2} & -{\sqrt{3}}/{2}
    \end{matrix} \right) \left( \begin{matrix}
        y_1 \\
        z_1 \\
        y_2 \\
        z_2
    \end{matrix} \right)\right|\\ 
    &\leq \frac{\varepsilon}{2}.
\end{split}
\end{align}
By combining this additional rotation by \( {5\pi}/{6} \) with intrinsic rigorous $\pi/2$ rotational symmetry of BCC lattice, we can delicately adjust the atomic positions in $\Lambda$ which have approximated 12-fold rotational symmetry, thereby ultimately obtaining a rigorously 12-fold symmetric PCPS $\Lambda'$. Furthermore, choosing a smaller $\varepsilon$ corresponds to applying a stricter probability threshold when extracting atomic positions from PFC results, thereby revealing a more pronounced rotational symmetry.

This theoretical framework also provides a natural explanation for the absence of 5- or 10-fold quasicrystal when $\theta = 36^\circ$, as well as for the occurrence (or lack thereof) of higher-order rotational symmetries when $\theta = 2{\pi}/{n},\ n\in \mathbb{N}$. Specifically, for $\theta = 36^\circ = 2{\pi}/{10}$, the projection of the 4 dimensional lattice $\Gamma$ into the internal space is given by
\begin{align}
     \pi_{\perp}(\Gamma) &= \left\{ 
 \left( \begin{matrix}
     1 & 0 & -\cos ({\pi}/{5}) & \sin ({\pi}/{5})  \\
    0 & 1 & -\sin ({\pi}/{5}) & -\cos ({\pi}/{5})  
 \end{matrix} \right)
  \mathbb{Z}^4 
 \right\} \nonumber\\
 &\nsupseteq \mathbb{Z}[\xi_{10}]
\end{align}
and thus does not generate 10-fold symmetry in both internal space and physical space. Moreover, for $\theta = 2{\pi}/{n}$, the Euler totient function $\phi$ satisfies $\phi(n) = 4$ if and only if $n \in \{5, 8, 10, 12\}$. This implies that only for these specific values of $n$ can the internal symmetry group be realized from a 4 dimensional lattice projection. Consequently, the BCC $[100]$ twist GBs are only capable of supporting quasicrystals of 8- and 12-fold.

\subsection{BCC $[110]$ tilt GB}
\label{sec:tilt}
We investigate $[110]$ symmetric tilt GBs in BCC crystals.
Two semi-infinite BCC bulk phases occupy the regions \(x < -L\) and \(x > L\), meeting at the \(x = 0\) plane.
Each crystal is rotated about the \(y\)-axis ($[110]$ direction) by tilt angles \(\pm \theta/2\) respectively. The resulting interfaces are periodic along the $y$-direction and consists of two alternating atomic layers exhibiting identical patterns, differing only by a translation. When $\varepsilon<a/2$, the CPS process in the PCPS framework enforces the relations $x_1=y_2,\ x_2 = y_1$ and $z_1=z_2$. Focusing on a single representative layer of atoms satisfying $x_i + y_i =0,\ i=1,2$, we introduce a coordinate transformation
\begin{align}
    h_i = \frac{x_i-y_i}{2},\quad i=1,2.
\end{align}
The 6 dimensional lattice $\Gamma$ is then reduced to a 2 dimensional lattice
\begin{align}
\Gamma := \left\{ 
a\begin{pmatrix}
2\sqrt{2}\cos (\theta/2) & -2\sin (\theta/2)\\
\sqrt{2}\sin (\theta/2) & \cos(\theta/2) 
\end{pmatrix}
\begin{pmatrix}
h \\
z
\end{pmatrix} 
\ \Big| \ \begin{pmatrix}
h \\
z
\end{pmatrix}   \in \mathbb{Z}^2
\right\}.
\end{align}
The interfacial pattern for this layer is described by the union of two CPS sets, constructed from the same 2 dimensional lattice $\Gamma$ but with distinct window
\begin{align}
    W_1&=\left[-\varepsilon,\varepsilon \right]\\ W_2 &= \left[u-\varepsilon,u+\varepsilon \right],\ u = \frac{\sqrt{2}\cos(\theta/2)-\sin(\theta/2)}{2}
\end{align}
and differing by a translation. This formulation is equivalent to the strip projection model introduced by Sutton in \cite{sutton1988irrational}. Including the $y$-$z$ plane, the Fourier-Bohr spectrum used in our numerical implementation is $\{\bm P\bm k\mid\bm k\in\mathbb{Z}^3\}$ where 
\begin{align}
    \bm  P=\frac{a^{-1}}{\sqrt{2}}\begin{pmatrix}
    1 & 1 & 0\\ 
    \sin (\theta/2) & -\sin(\theta/2) & \sqrt{2}\cos(\theta/2)
    \end{pmatrix}.
\end{align}
The LB model parameters are also set to $\alpha = 0.0,\ \gamma=0.5$, consistent with the stability regime of the BCC phase.

\subsubsection{Periodic tilt GBs}

\begin{figure}[!t]
	\centering
	\includegraphics[width=0.3\textwidth]{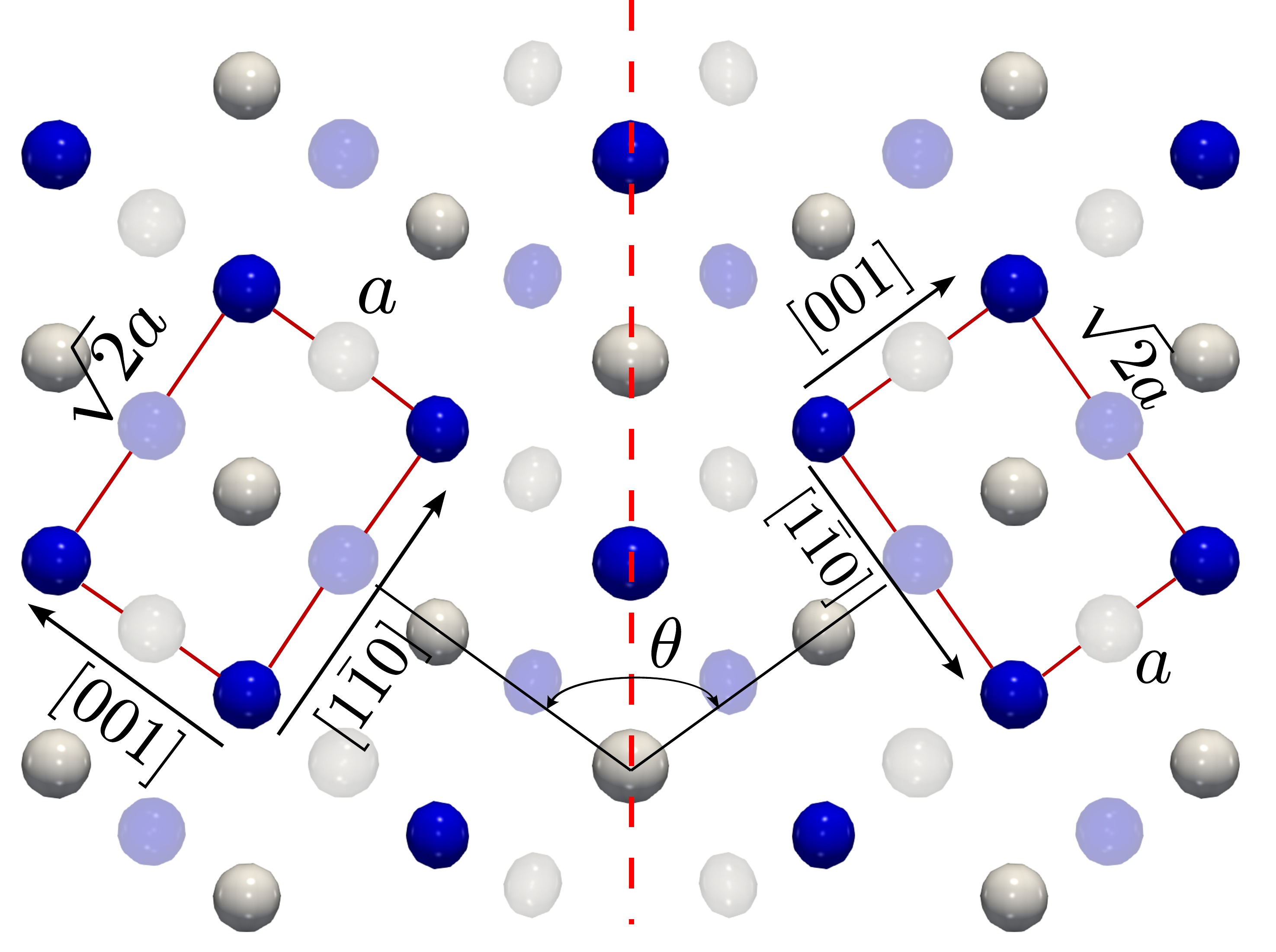}
	\caption{Twin boundary in BCC with the misorientation $\theta =2 \arctan \sqrt{2}$.
	Adjacent $(220)$ atomic layers are distinguished by staining them as transparent and opaque.  
    Blue spheres represent corner atoms of the BCC crystal, while white spheres represent body-centered atoms.}
	\label{fig:TGBperiodic}
\end{figure}

Similar to the twist GB cases, BCC $[110]$ symmetric tilt GBs can be categorized as either periodic or quasiperiodic. When $\sqrt{2}\tan (\theta/2)\in \mathbb{Q}$, the two bulk lattices with misorientation $\theta$ periodically coincide at the interface, leading to a periodic GB. Otherwise, the GB is quasiperiodic.
We take the twin boundary with misorientation $\theta = 2\arctan \sqrt{2}$ as an example of periodic BCC symmetric tilt GBs.
Numerical results are shown in \cref{fig:TGBperiodic}.
Atomic positions are extracted from the $70$\% thresholded level set of the density field. Atoms at the interface coincide commensurately at the positions predicted by the CSL theory, forming a periodic pattern.

\subsubsection{Quasiperiodic tilt GBs}
\begin{figure*}[!t]
	\centering
	\subfigure[$\theta = 90^\circ$.]{
	\includegraphics[width=0.6\textwidth]{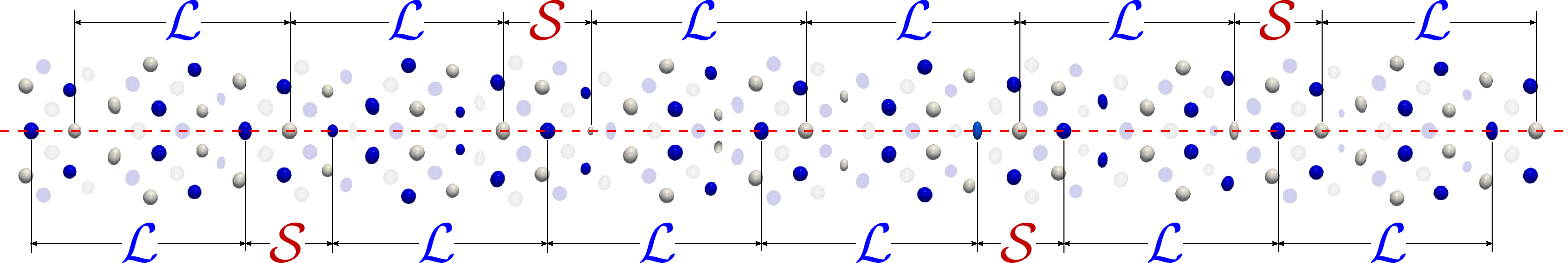}}
	\subfigure[$\theta = 2 \arctan ((\sqrt{3}+1)/\sqrt{2})$.]{
	\includegraphics[width=0.8\textwidth]{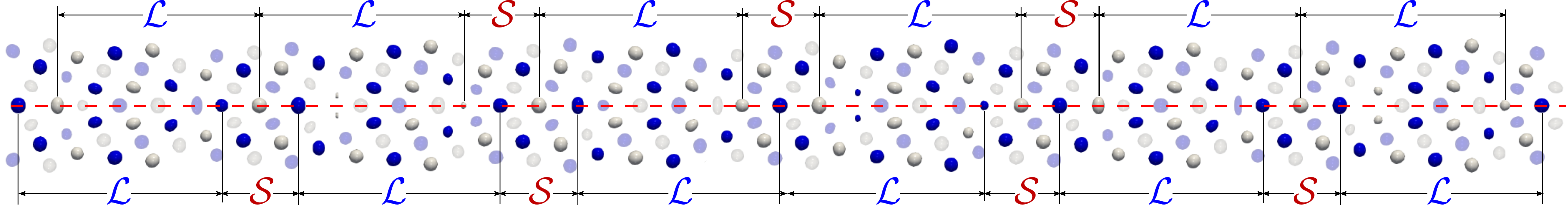}}
	\caption{BCC $[110]$ symmetric tilt GBs at (a) $\theta = 90^\circ$; (b) $\theta = 2\arctan ((\sqrt{3}+1)/\sqrt{2})$. GB pattern consists of 2 types of spacing $\mathcal{L}$ and $\mathcal{S}$. The ratio $\mathcal{L}/\mathcal{S}$ is $\mathcal{L}/\mathcal{S} = \sqrt{2}+1$ in (a)  and $\mathcal{L}/\mathcal{S} = \sqrt{3}+1$ in (b).}
	\label{fig:ttgb}
\end{figure*}

For most misorientation angles $\theta$, BCC symmetric tilt GBs are quasiperiodic according to the given PCPS model. We present numerical results for BCC $[110]$ tilt GBs at tilt angles $\theta = 90^\circ$ and $\theta = 2\arctan((\sqrt{3}+1)/\sqrt{2})$ in \cref{fig:ttgb}.

On the GB plane, blue and white spheres within the same atomic layer exhibit two atomic spacings, corresponding to the ``$\mathcal{L}$" (long spacing) and ``$\mathcal{S}$" (short spacing). Taking the $\theta = 90^\circ$ case as an example, in this case their lengths are $\mathcal{L} = (2\sqrt{2} + 3)a/2$ and $\mathcal{S} = (\sqrt{2} + 1)a/2$, with a ratio $\mathcal{L}/\mathcal{S} = \sqrt{2} + 1$.  
The arrangement of atoms at the GB plane follows a two-letter substitution sequence, which is a representative feature of quasiperiodicity. This quasiperiodic sequence can be generated by the following substitution rule 
\begin{align}
\begin{cases}
\mathcal{L}\mapsto \mathcal{LSL}\\
\mathcal{S}\mapsto \mathcal{L}
\end{cases}
\end{align}
and correspond to one of the fixed points of this substitutional dynamical system. Since all fixed points associated with a given substitution rule are locally indistinguishable (see \cite{Baake2002}), and only finite patches of the GB structure are accessible in visual observations, it is unnecessary to distinguish between them. 

The case of $\theta = 2\arctan((\sqrt{3}+1)/\sqrt{2})$ shows similar quasiperiodic features with lengths $\mathcal{L} = (3\sqrt{6} + 5\sqrt{2})/\sqrt{6 - 2 \sqrt{3}}a,\ \mathcal{S} = (\sqrt{6} + 2\sqrt{2})/\sqrt{6 - 2 \sqrt{3}}a$ and ratio $\mathcal{L}/\mathcal{S} = \sqrt{3}+1$. For other misorientations, GB patterns can also be generated by substitutional dynamical systems, which may involve three characteristic spacings ``$\mathcal{L}$" (long spacing), ``$\mathcal{M}$" (middle spacing) and ``$\mathcal{S}$" (short spacing). The emergence of this quasiperiodic arrangement can be attributed to the intrinsic mathematical connection between CPS constructed on quadratic irrational lattices and the existence of corresponding substitution rules, see \cite{masakova2000substitution}. This relationship, together with our PCPS theory, provides a natural theoretical explanation for the appearance of the observed quasiperiodicity in tilt GB structures.

\subsection{BCC/FCC Interfaces}
\label{sec:interface}
Interfaces between BCC and FCC bulk phases with different orientations are investigated in this section. The BCC bulk phase occupying $x < -L$ is rotated about the $x$-axis by a twist angle $\theta$, while the FCC bulk phase occupying $x > L$ remains unrotated. Two bulk phases possess distinct lattice constants, denoted by $a$ for BCC and $b$ for FCC. The restriction of the FCC bulk lattice to the interfacial plane can be written as the union $b\mathbb{Z}^2 \cup (b\mathbb{Z}^2 +\bm u)$, where $\bm u = (b/2,b/2)^T$. As a result, the standard PCPS can be reduced to the union of two CPS sets projected from a 4 dimensional lattice to $\mathbb{R}^2$, similar to the construction described in \cref{sec:twist}. These two CPS sets share the same 4 dimensional lattice
\begin{align}
\Gamma := \left\{ 
\begin{pmatrix}
a\bm{R}_\theta \quad \ -b\bm{I} \\
\ a\bm{R}_\theta/2 \quad \ \ b\bm{I}/2
\end{pmatrix}
\begin{pmatrix}
\bm{x} \\
\bm{y}
\end{pmatrix} 
\ \Big| \ \begin{pmatrix}
\bm{x} \\
\bm{y}
\end{pmatrix}  \in \mathbb{Z}^4 
\right\},
\end{align}
with distinct cut windows $W_1 = \overline{B(\bm 0,\varepsilon)}$ and $W_2 = \overline{B(\bm u,\varepsilon)}$. Parameters in LB calculation are chosen as $\alpha=0.0,\gamma=1.55$ to ensure compatibility with both the BCC and FCC phases. The lattice constants for two bulk phases are then computed as $a\approx 9.26$ and $b\approx 11.75$. According to the PCPS framework, a periodic interface requires $b^2 / a^2 \in \mathbb{Q}$. Although this ratio is rational in numerical computation due to finite numerical precision, the associated $\Sigma$ value is prohibitively large. Hence, the BCC/FCC interface is treated as quasiperiodic for any twist angle $\theta$.

\begin{figure*}[!t]
	\centering
	\includegraphics[width=0.05\textwidth]{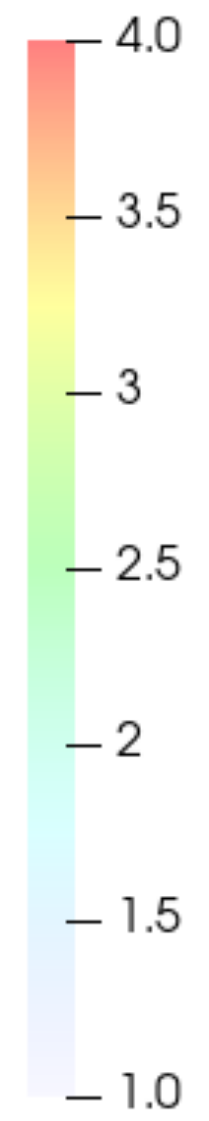}
	\subfigure[$\theta = 45^\circ$]{
	\includegraphics[width=0.30\textwidth]{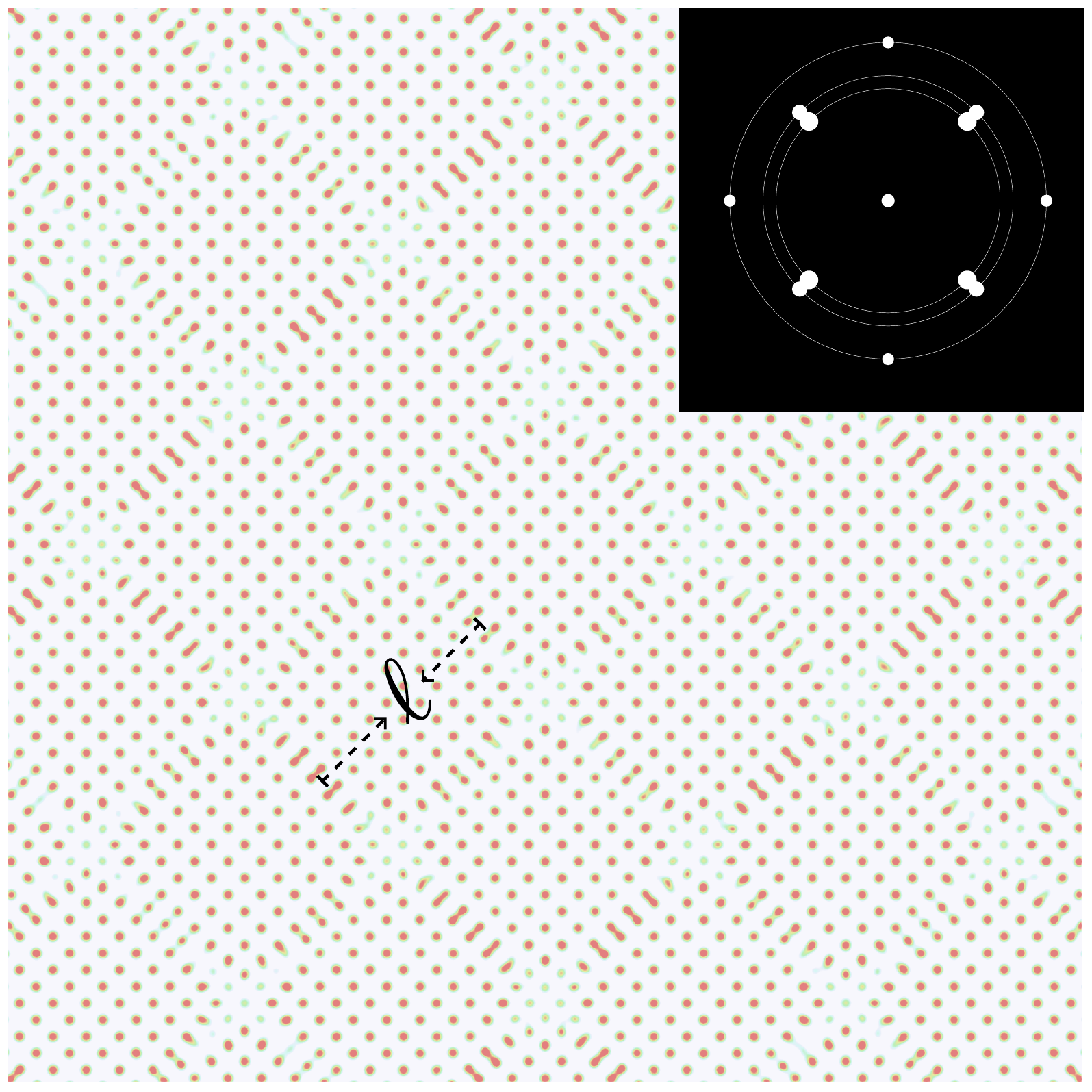}}
	\subfigure[$\theta = 40^\circ$]{
	\includegraphics[width=0.30\textwidth]{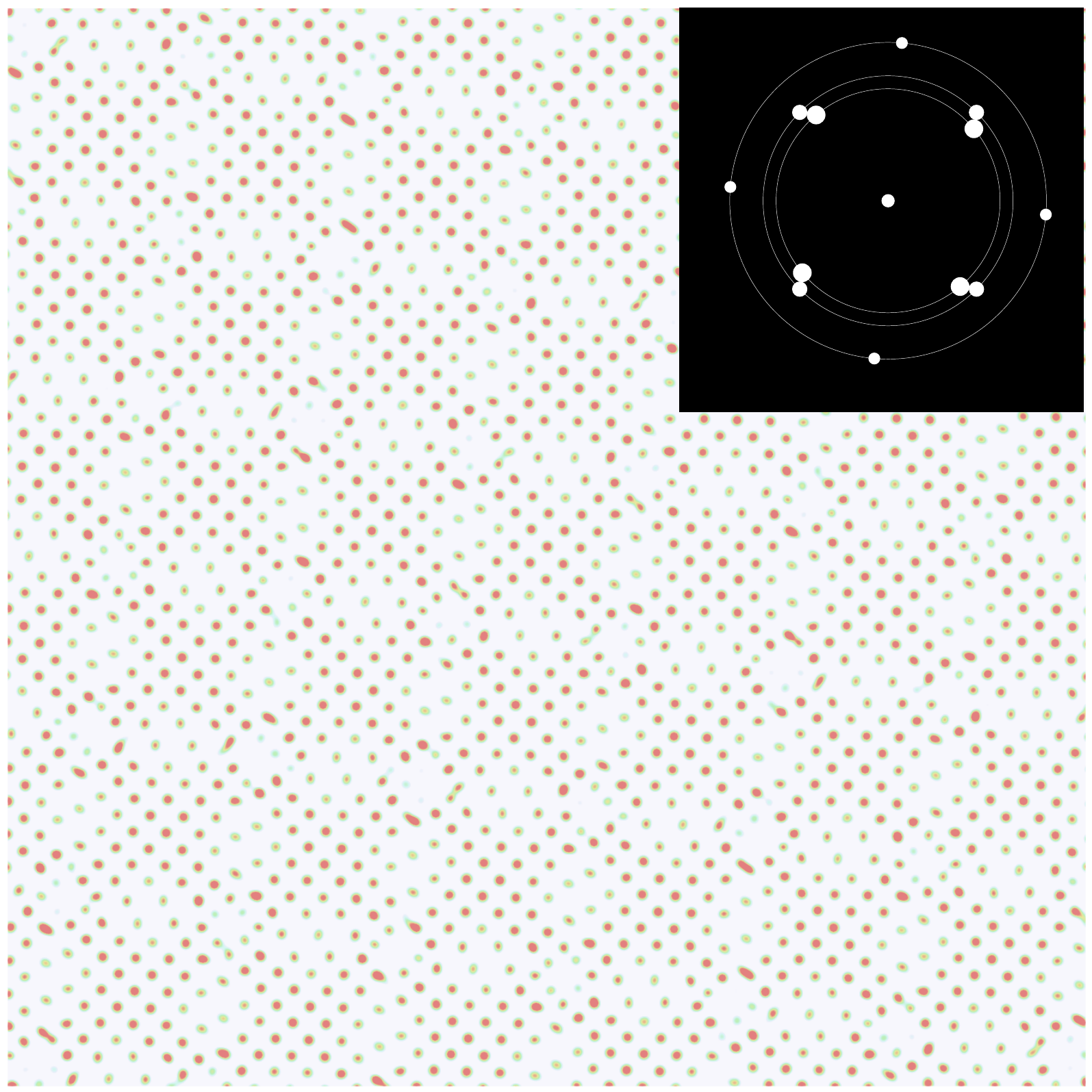}}
	\subfigure[$\theta = 35^\circ$]{
	\includegraphics[width=0.30\textwidth]{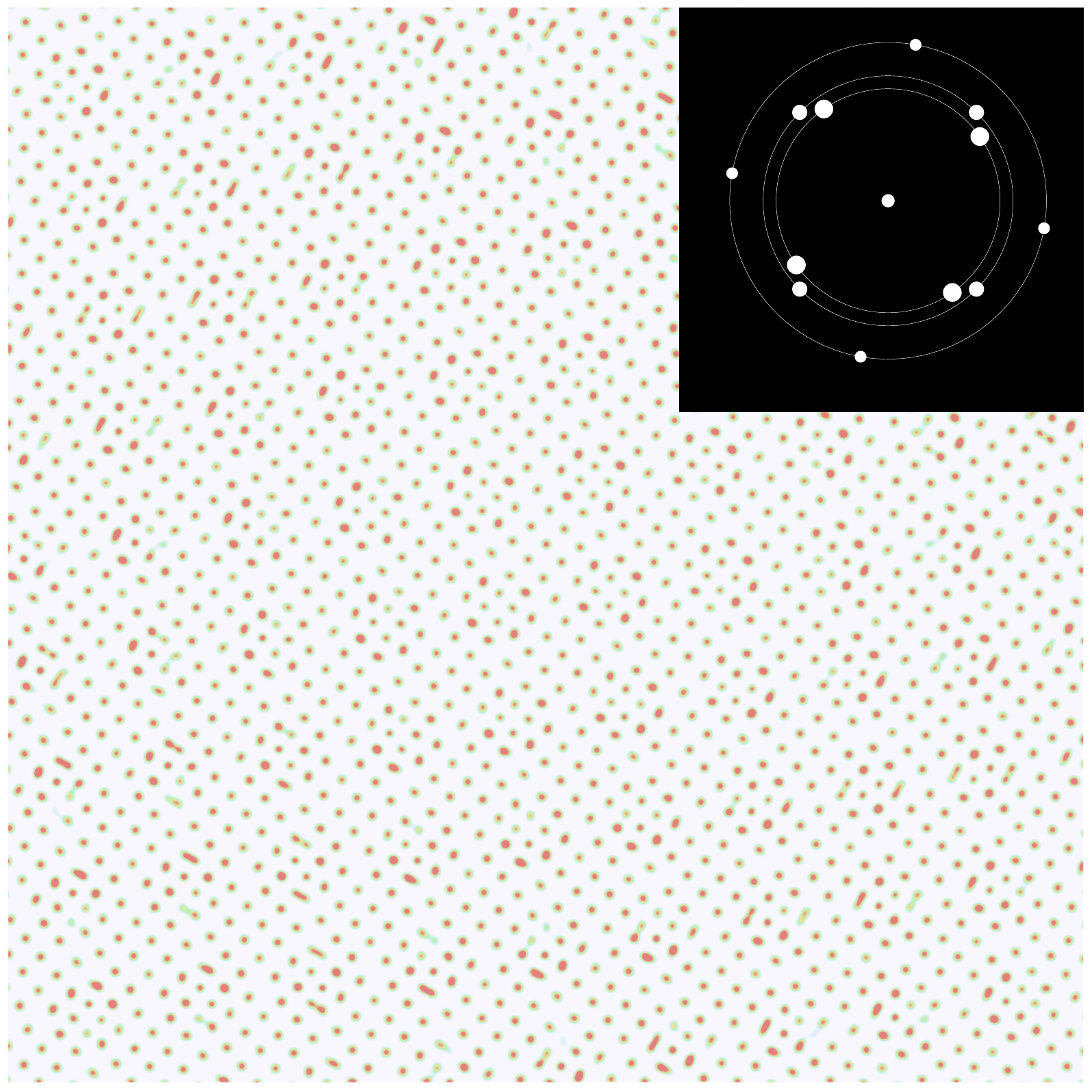}}
	\caption{Low-angle BCC/FCC interfaces with $(100)$ contact plane and twist angles of (a) $\theta = 45^\circ$; (b) $\theta =40^\circ$; and (c) $\theta = 35^\circ$.}
	\label{fig:low_angle_interface}
\end{figure*}

\begin{figure*}[!t]
	\centering
	\subfigure[$\theta = 0^\circ$]{
	\includegraphics[width=0.30\textwidth]{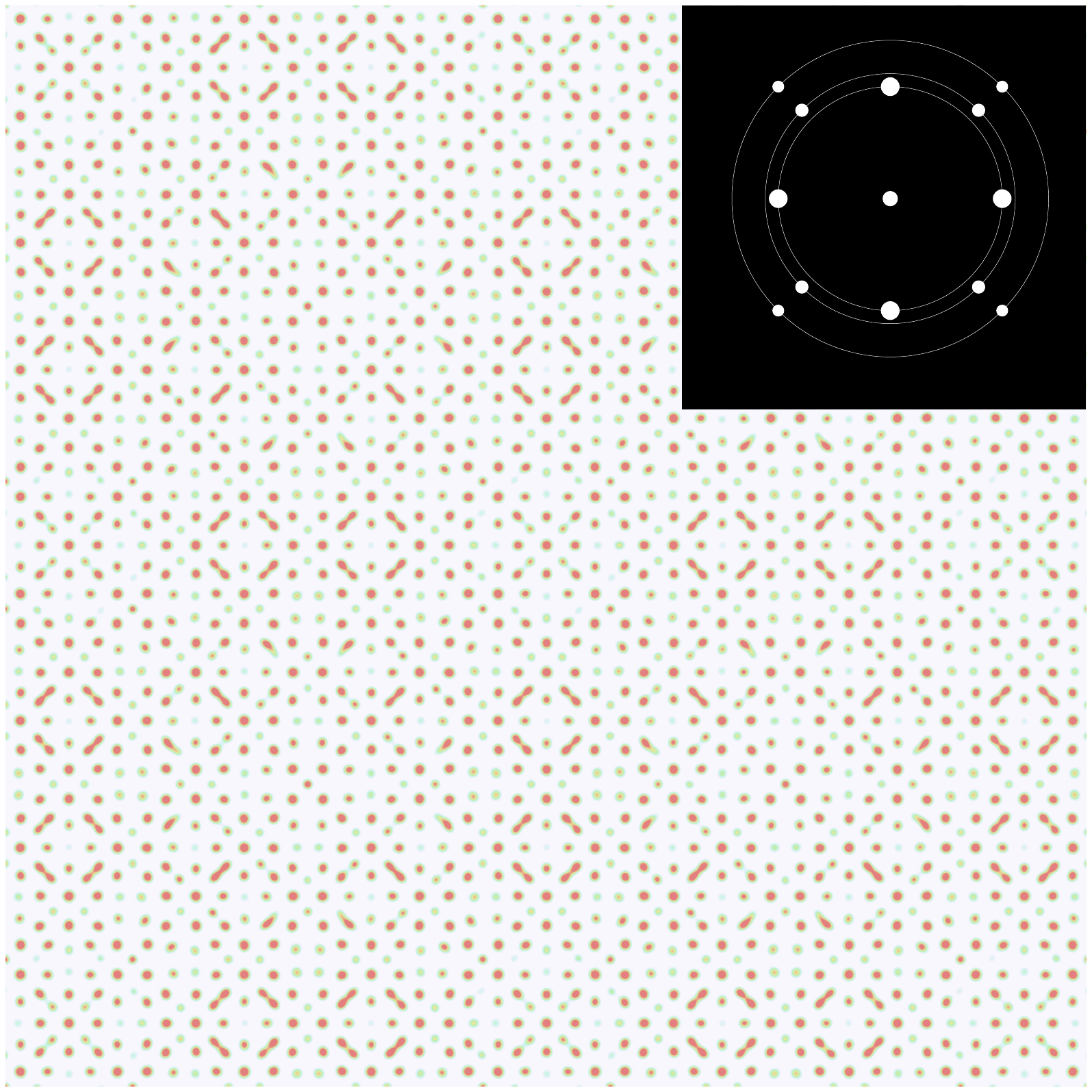}}
	\subfigure[$\theta = 15^\circ$]{
	\includegraphics[width=0.30\textwidth]{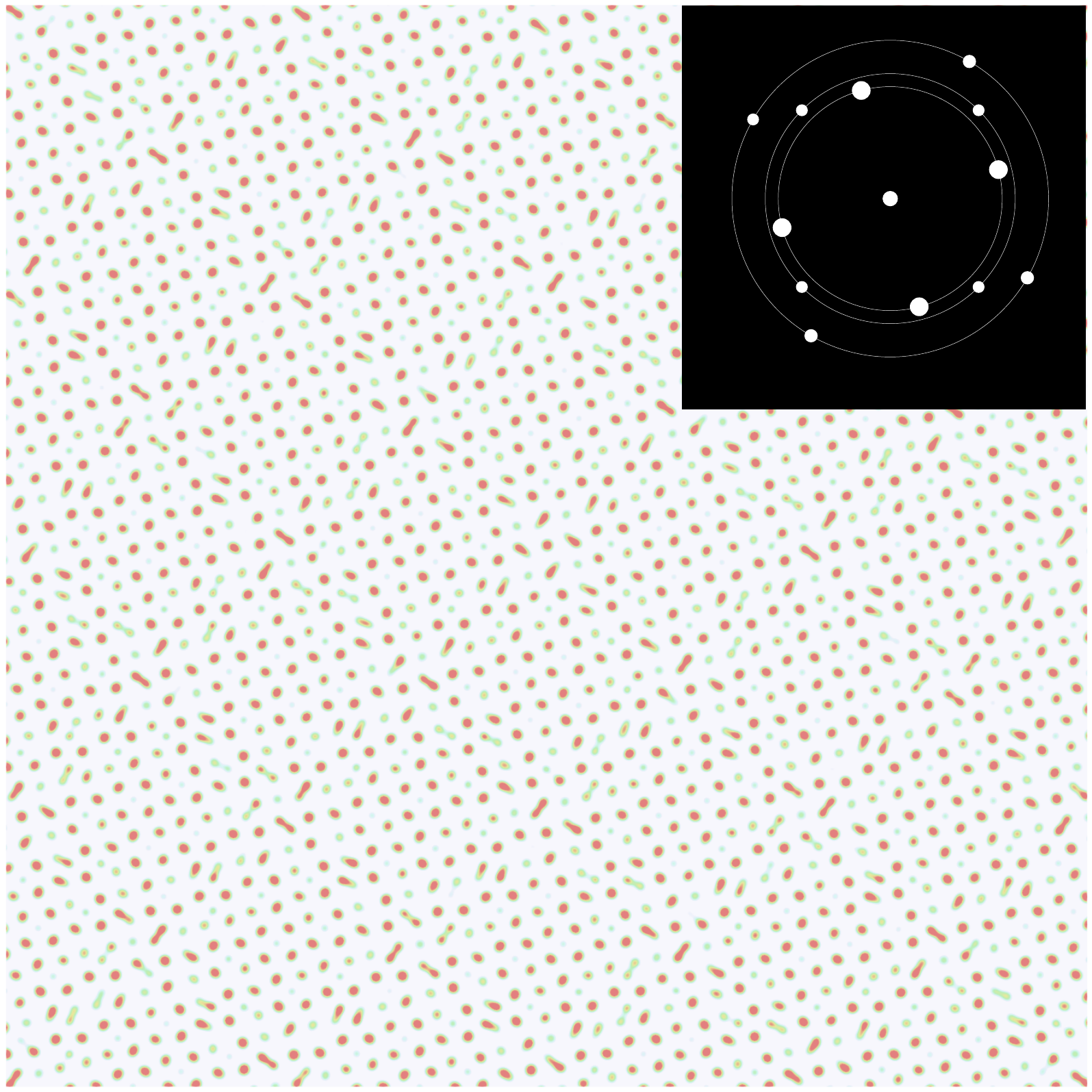}}
	\subfigure[$\theta = 30^\circ$]{
	\includegraphics[width=0.30\textwidth]{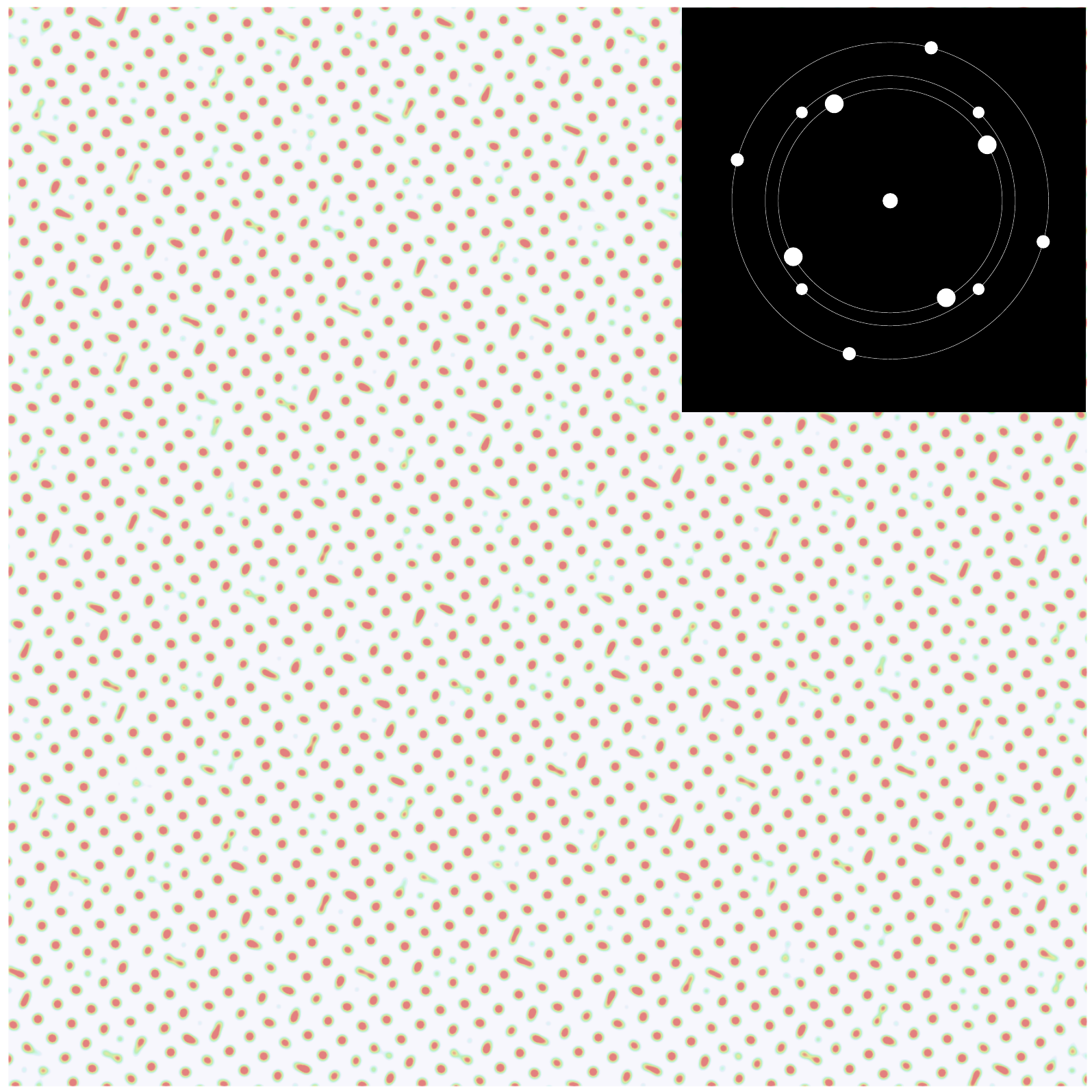}}
	\caption{High-angle BCC/FCC interfaces with $(100)$ contact plane and twist angles of (a) $\theta = 0^\circ$; (b) $\theta =15^\circ$; and (c) $\theta = 30^\circ$.}
	\label{fig:high_angle_interface}
\end{figure*}

\paragraph{Low-angle interfaces}
Parallel to the interface, the FCC bulk lattice can be viewed as a planar square lattice with primitive vectors $(b/\sqrt{2}, \pm b/\sqrt{2})^{\mathrm{T}}$. Therefore, when the misorientation angle $\theta$ approaches $45^\circ$, our PCPS theory predicts the emergence of dislocation networks similar to those observed in BCC low-angle twist GBs. Accordingly, we categorize cases satisfying $|45^\circ - \theta|\lesssim 15^\circ$ as low-angle interfaces. Analogous the BCC low-angle twist GB cases, the side length $\ell$ of these dislocation networks can be estimated by computing the spatial accommodation required between adjacent grains as
\begin{align}
    \left|\frac{\ell }{\mathrm{min} \{a,b/\sqrt{2}\}\cos\theta} - \frac{\ell }{\mathrm{max} \{a,b/\sqrt{2}\}}\right| \approx 1. 
\end{align}
Hence we can deduce that 
\begin{align}
    \ell \approx \frac{1}{\sqrt{2}}\frac{ab\cos\theta}{\mathrm{max} \{a,b/\sqrt{2}\} - \mathrm{min} \{a,b/\sqrt{2}\}\cos\theta}.
\end{align}
We present out numerical results for BCC/FCC low-angle interfaces at $\theta = 45^\circ$, $40^\circ$ and $35^\circ$ in \cref{fig:low_angle_interface}.
At $\theta = 45^\circ$, the dislocation network exhibits a square-like pattern. As $\theta$ decreases, the characteristic size of the dislocation network correspondingly decreases, which becomes progressively distorted and eventually disordered. The Fourier-Bohr spectrum of resulting PCPS is then given by $\{\bm P \bm k \mid \bm k \in \mathbb{Z}^4\},\ \bm P = (a^{-1}\bm R_{-\theta}, b^{-1}\bm I)$.
The first and second level spectral components are all correspond to \cref{tab:interface_spectra}.
\begin{table}[!htbp]
    \centering
    \caption{Spectral components of the first and second levels of intensity in BCC/FCC interfaces.}
    \label{tab:interface_spectra}
    \begin{tabular}{cc}
        \toprule
        ~~Intensity~~ & Spectra $\bm P \bm k$ \\
        \midrule
        1st level & $\bm{P}(\pm 1,0,0,0)^T,\ \bm{P}(0,\pm 1,0,0)^T$ \\
        \\[-0.5em]
        2nd level & $\bm{P}(0,0,\pm 1,\pm 1)^T$ \\
        \bottomrule
    \end{tabular}
\end{table}

\begin{figure}[!hpbt]
	\centering
	\includegraphics[width=0.4\textwidth]{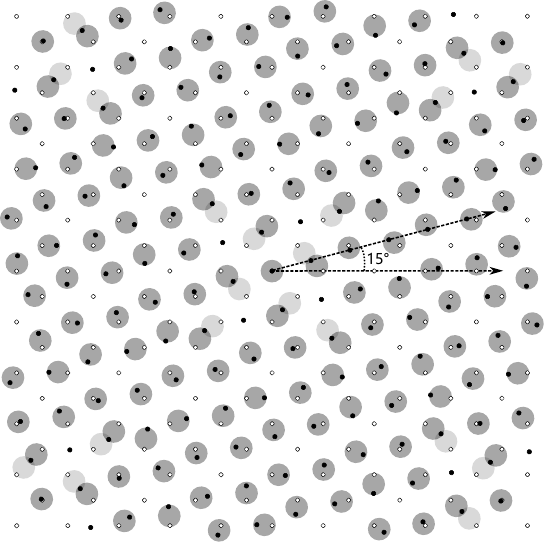}
	\caption{The PCPS diagram for the BCC/FCC interface at $\theta = 15^\circ$. The tolerance parameter is set to $\varepsilon \approx 4.754$ to match the 70\% thresholded level-set density. For pairs of PCPS points that still remain sufficiently close, those generated from proximal pairs with larger separations are shown in light grey, while those from smaller separations are rendered normally.
	}
	\label{fig:interfaceGCSN}
\end{figure}

\paragraph{High-angle interfaces}
When $|45^\circ - \theta| \gtrsim 15^\circ $, the dislocation networks observed in low-angle cases become completely disrupted, giving rise to an apparently quasiperiodic interfacial structure. We presents the numerical results for the BCC/FCC interfaces at twist angles $\theta = 0^\circ,\ 15^\circ,\ 30^\circ$ in \cref{fig:high_angle_interface}, along with the PCPS diagram for $\theta = 15^\circ$ in \cref{fig:interfaceGCSN}. As observed, certain pairs of points in PCPS diagram still remain sufficiently close to each other. Since the PCPS framework allows an arbitrary displacement of up to $\varepsilon/2$ for each atom, it is physically reasonable to consider additional atomic merging for such pairs. In numerical simulations, the locations of these pairs coincide with regions that host droplet-like local density distributions. PCPS points generated from proximal pairs with larger separations correspond to the tail regions of these local distributions, while normally rendered points correspond to their denser heads. By comparing these results, it is evident that the PCPS model continues to provide accurate predictions of the thresholded level sets, effectively capturing the spatial distribution of interfacial atomic densities.

The transition from low-angle to high-angle configurations, accompanied by the disappearance of dislocation networks, resembles the breakdown observed in high-angle BCC twist GB cases. This phenomenon admits a similar explanation within the PCPS framework, as arising from the increasing spatial separation between adjacent grains. Given the generality of this mechanism, we infer that such breakdown behavior constitutes a universal feature of 2 dimensional quasiperiodic interfaces, which is also observed in other types of interfaces \cite{feng2015energy}.

\section{Conclusions and outlook}
\label{sec:conclusion}
In this work, we establish a unified quasiperiodicity modelling framework for interfaces, addressing fundamental challenges in the structural configuration of incommensurate systems. Leveraging the developed PCPS theory, we demonstrate that interfaces can be modeled as rigorous cut-and-project sets, thereby providing a systematic route for capturing quasiperiodicity in interfacial structures. This theoretical foundation directly motivates the implementation of projection method within the LB model, ensuring both accuracy and reliability in long-range simulations. Within this framework, we fully characterize several representative interfacial structures, including BCC $[100]$ twist GBs, BCC $[110]$ tilt GB and BCC/FCC interfaces, along with their associated pattern formations across arbitrary misorientations.
 
Our computations reveal that interfacial structures and structural patterns vary systematically with the choice of bulk phases. The distribution of high-density regions, represented by thresholded level sets, shows excellent agreement with predictions from the PCPS model. Quasiperiodic features such as spectral characteristics, FLC and repetitivity are verified in the simulated interfaces. For BCC $[100]$ twist GBs and BCC/FCC interfaces, the resulting interfacial structures can be categorized into low-angle and high-angle cases. Low-angle interfaces exhibit well-ordered dislocation networks with predictable side lengths, whereas high-angle interfaces demonstrates distinct quasiperiodic ordering. Quasicrystals can also emerge in high-angle cases as observed in BCC $[100]$ twist GBs at \( \theta = 30^\circ \) and \( 45^\circ \), corresponding to visually evident 12- and 8-fold rotational symmetries respectively. These features arise naturally from the internal space symmetry of the underlying PCPS construction. For BCC $[110]$ tilt GBs, our theoretical model provides a rigorous explanation for the appearance of quasiperiodically distributed interfacial sequences. Other quasiperiodic features, such as self-similar scaling ratios, can also be deduced from PCPS and verified by our numerical framework. 
 
Overall, our combined theoretical and computational framework applies not only to the interfacial structures studied here but also to a broad class of incommensurate systems. In the case of BCC $[100]$ twist GBs, our model further predicts the absence of other quasicrystals such as 5- and 10-fold, which is attributed to both the lack of corresponding symmetry in the internal space and the arithmetic constraint \( \phi(n) = 4 \) on the Euler totient function. From both a physical standpoint and a materials design perspective, our analysis indicates that realizing $n$-fold quasicrystals with $\phi(n) > 4$ requires a higher-dimensional CPS capable of generating the ring $\mathbb{Z}[\xi_n]$ in the internal space. This insight, combined with further development in PCPS theoretical model and our numerical computational method, opens new avenues for engineering quasicrystals and quasiperiodic materials.

\bibliography{ref}

@PREAMBLE{
 "\providecommand{\noopsort}[1]{}" 
 # "\providecommand{\singleletter}[1]{#1}%" 
}

@article{li2015defects,
  title={Defects in the self-assembly of block copolymers and their relevance for directed self-assembly},
  author={Li, Weihua and M{\"u}ller, Marcus},
  journal={Annu. Rev. Chem. Biomol.},
  volume={6},
  number={1},
  pages={187--216},
  year={2015},
  publisher={Annual Reviews}
}

@article{chimi2001accumulation,
  title={Accumulation and recovery of defects in ion-irradiated nanocrystalline gold},
  author={Chimi, Y and Iwase, A and Ishikawa, N and Kobiyama, M and Inami, T and Okuda, S},
  journal={J. Nucl. Mater.},
  volume={297},
  number={3},
  pages={355--357},
  year={2001},
  publisher={Elsevier}
}

@article{ogata2005energy,
  title={Energy landscape of deformation twinning in bcc and fcc metals},
  author={Ogata, Shigenobu and Li, Ju and Yip, Sidney},
  journal={Phys. Rev. B},
  volume={71},
  number={22},
  pages={224102},
  year={2005},
  publisher={APS}
}

@article{xiao2016effect,
  title={Effect of grain boundary on the mechanical behaviors of irradiated metals: a review},
  author={Xiao, XiaZi and Chu, HaiJian and Duan, HuiLing},
  journal={Sci. China-Phys. Mech. Astron.},
  volume={59},
  pages={1--11},
  year={2016},
  publisher={Springer}
}

@article{lanccon2019incommensurate,
  title={Incommensurate grain boundary in silicon and the silver-ratio sequence},
  author={Lan{\c{c}}on, Fr{\'e}d{\'e}ric and Gunkelmann, Nina and Caliste, Damien and Rouvi{\`e}re, Jean-Luc},
  journal={Phys. Rev. B},
  volume={100},
  number={11},
  pages={115307},
  year={2019},
  publisher={APS}
}

@article{yamanaka2017phase,
  title={Phase field crystal simulation of grain boundary motion, grain rotation and dislocation reactions in a BCC bicrystal},
  author={Yamanaka, Akinori and McReynolds, Kevin and Voorhees, Peter W},
  journal={Acta Mater.},
  volume={133},
  pages={160--171},
  year={2017},
  publisher={Elsevier}
}

@Inbook{Baake2002,
author={Baake, Michael},
title={A Guide to Mathematical Quasicrystals},
bookTitle={Quasicrystals: An Introduction to Structure, Physical Properties and Applications},
year={2002},
pages={17--48},
isbn={978-3-662-05028-6}
}

@article{seki2023incommensurate,
  title={Incommensurate grain-boundary atomic structure},
  author={Seki, Takehito and Futazuka, Toshihiro and Morishige, Nobusato and Matsubara, Ryo and Ikuhara, Yuichi and Shibata, Naoya},
  journal={Nat. Commun.},
  volume={14},
  number={1},
  pages={7806},
  year={2023},
  publisher={Nature Publishing Group UK London}
}

@book{baake2013aperiodic,
  title={Aperiodic order},
  author={Baake, Michael and Grimm, Uwe},
  volume={1},
  year={2013},
  publisher={Cambridge University Press}
}

@article{jiang2014numerical,
	title = {Numerical methods for quasicrystals},
	volume = {256},
	issn = {00219991},
	urldate = {2022-07-25},
	journal = {J. Comput. Phys.},
	author = {Jiang, Kai and Zhang, Pingwen},
	month = jan,
	year = {2014},
	pages = {428--440}
}

@article{trevino2023aperiodic,
  title={Aperiodic tilings, order, and randomness},
  author={Trevi{\~n}o, Rodrigo},
  journal={Not. Am. Math. Soc.},
  volume={70},
  number={8},
  pages={1179--1191},
  year={2023}
}

@book{meyer1972algebraic,
  title={Algebraic numbers and harmonic analysis},
  author={Meyer, Yves},
  volume={2},
  year={1972},
  publisher={Elsevier}
}

@book{randle2024role,
  title={The role of the coincidence site lattice in grain boundary engineering},
  author={Randle, Valerie},
  year={2024},
  publisher={CRC Press}
}

@article{GCSNAra,
author = {Aragón, José and Romeu, Luis and Beltrán, L. and Gómez-Rodríguez, Alfredo},
year = {1997},
month = {11},
pages = {772-780},
title = {Grain Boundaries as Projections from Higher-Dimensional Lattices},
volume = {53},
journal = {Acta Crystallogr. A},
doi = {10.1107/S010876739700737X}
}

@article{zhang1993lattice,
  title={O-lattice analyses of interfacial misfit. {I}. {G}eneral considerations},
  author={Zhang, W-Z and Purdy, GR},
  journal={Philos. Mag. A},
  volume={68},
  number={2},
  pages={279--290},
  year={1993},
  publisher={Taylor \& Francis}
}

@article{fu2020optical,
  title={Optical soliton formation controlled by angle twisting in photonic moir{\'e} lattices},
  author={Fu, Qidong and Wang, Peng and Huang, Changming and Kartashov, Yaroslav V and Torner, Lluis and Konotop, Vladimir V and Ye, Fangwei},
  journal={Nat. Photonics},
  volume={14},
  number={11},
  pages={663--668},
  year={2020},
  publisher={Nature Publishing Group UK London}
}

@article{frolov2018grain,
  title={Grain boundary phases in bcc metals},
  author={Frolov, Timofey and Setyawan, Wahyu and Kurtz, RJ and Marian, Jaime and Oganov, Artem R and Rudd, Robert E and Zhu, Qiang},
  journal={Nanoscale},
  volume={10},
  number={17},
  pages={8253--8268},
  year={2018},
  publisher={Royal Society of Chemistry}
}

@article{wang2018grain,
  title={Grain boundaries in bcc-Fe: A density-functional theory and tight-binding study},
  author={Wang, Jingliang and Madsen, Georg KH and Drautz, Ralf},
  journal={Model. Simul. Mater. Sci. Eng.},
  volume={26},
  number={2},
  pages={025008},
  year={2018},
  publisher={IOP Publishing}
}

@article{yin2021transition,
  title={Transition pathways connecting crystals and quasicrystals},
  author={Yin, Jianyuan and Jiang, Kai and Shi, An-Chang and Zhang, Pingwen and Zhang, Lei},
  journal={Proc. Nat. Acad. Sci.},
  volume={118},
  number={49},
  pages={e2106230118},
  year={2021},
  publisher={National Acad Sciences}
}

@article{vitek1968intrinsic,
  title={Intrinsic stacking faults in body-centred cubic crystals},
  author={Vitek, Vaclav},
  journal={Philos. Mag.},
  volume={18},
  number={154},
  pages={773--786},
  year={1968},
  publisher={Taylor \& Francis}
}

@article{vitek1980structure,
  title={Structure of tilt grain boundaries in bcc metals},
  author={Vitek, V and Smith, DA and Pond, RC},
  journal={Philos. Mag. A},
  volume={41},
  number={5},
  pages={649--663},
  year={1980},
  publisher={Taylor \& Francis}
}

@article{xu2017computing,
  title={Computing optimal interfacial structure of modulated phases},
  author={Xu, Jie and Wang, Chu and Shi, An-Chang and Zhang, Pingwen},
  journal={Commun. Comput. Phys.},
  volume={21},
  number={1},
  pages={1--15},
  year={2017},
  publisher={Cambridge University Press}
}

@article{cao2021computing,
  title={Computing interface with quasiperiodicity},
  author={Cao, Duo and Shen, Jie and Xu, Jie},
  journal={J. Comput. Phys.},
  volume={424},
  pages={109863},
  year={2021},
  publisher={Elsevier}
}

@article{brazovskii1975phase,
  title={Phase transition of an isotropic system to a nonuniform state},
  author={Brazovskii, Sov},
  journal={Sov. Phys. JETP},
  volume={41},
  pages={85},
  year={1975},
  note = {[Zh. Eksp. Teor. Fiz. \textbf{68}, 175 (1975)]}
}

@article{zhang2008efficient,
  title={An efficient numerical method of {L}andau-{B}razovskii model},
  author={Zhang, Pingwen and Zhang, Xinwei},
  journal={J. Comput. Phys.},
  volume={227},
  number={11},
  pages={5859--5870},
  year={2008},
  publisher={Elsevier}
}

@article{fredrickson1987fluctuation,
  title={Fluctuation effects in the theory of microphase separation in block copolymers},
  author={Fredrickson, Glenn H and Helfand, Eugene},
  journal={J. Chem. Phys.},
  volume={87},
  number={1},
  pages={697--705},
  year={1987},
  publisher={American Institute of Physics}
}

@article{leibler1980theory,
  title={Theory of microphase separation in block copolymers},
  author={Leibler, Ludwik},
  journal={Macromolecules},
  volume={13},
  number={6},
  pages={1602--1617},
  year={1980},
  publisher={ACS Publications}
}

@phdthesis{mcclenagan2019landau,
  title={Landau theory of complex ordered phases},
  author={McClenagan, Duncan},
  year={2019},
  school = {McMaster University, Hamilton, ON, Canada}
}

@inproceedings{jiang2018numerical,
  title={Numerical mathematics of quasicrystals},
  author={Jiang, Kai and Zhang, Pingwen},
  booktitle={Proc. Int. Cong. of Math., Vol. 3},
  pages={3575--3594},
  year={2018},
}

@book{ShenJiebook,
author = {Shen, Jie and Tang, Tao and Wang, Li-Lian},
year = {2011},
month = {01},
title = {Spectral Mathod: Algorithms, Analysis and Applications},
volume = {41},
isbn = {978-3-540-71040-0},
journal = {Springer Ser. Comput. Math.},
}

@book{washington2012introduction,
  title={Introduction to cyclotomic fields},
  author={Washington, Lawrence C},
  volume={83},
  year={2012},
  publisher={Springer Science \& Business Media}
}

@article{feng2015energy,
  title={The energy and structure of (110) twist grain boundary in tungsten},
  author={Feng, Yaxin and Shang, Jiaxiang and Liu, Zenghui and Lu, Guanghong},
  journal={Appl. Surf. Sci.},
  volume={357},
  pages={262--267},
  year={2015},
  publisher={Elsevier}
}

@article{schwartz1985atomic,
  title={Atomic structure of (001) twist boundaries in fcc metals Structural unit model},
  author={Schwartz, D and Vitek, V and Sutton, AP},
  journal={Philos. Mag. A},
  volume={51},
  number={4},
  pages={499--520},
  year={1985},
  publisher={Taylor \& Francis}
}

@article{schober1970quantitative,
  title={Quantitative observation of misfit dislocation arrays in low and high angle twist grain boundaries},
  author={Schober, T and Balluffi, RW},
  journal={Philos. Mag.},
  volume={21},
  number={169},
  pages={109--123},
  year={1970},
  publisher={Taylor \& Francis}
}

@article{dai2014atomistic,
  title={Atomistic, generalized {P}eierls--{N}abarro and analytical models for (111) twist boundaries in {A}l, {C}u and {N}i for all twist angles},
  author={Dai, Shuyang and Xiang, Yang and Srolovitz, David J},
  journal={Acta Mater.},
  volume={69},
  pages={162--174},
  year={2014},
  publisher={Elsevier}
}

@article{yang2010atomic,
  title={Atomic scale modeling of $\{$110$\}$ twist grain boundaries in $\alpha$-iron: Structure and energy properties},
  author={Yang, JB and Nagai, Y and Hasegawa, M and Osetsky, Yu N},
  journal={Philos. Mag.},
  volume={90},
  number={7-8},
  pages={991--1000},
  year={2010},
  publisher={Taylor \& Francis}
}

@article{zou2025quasiperiodic,
title = {Quasiperiodic [110] symmetric tilt {FCC} grain boundaries},
journal = {Comput. Mater. Sci.},
volume = {253},
pages = {113811},
year = {2025},
issn = {0927-0256},
doi = {https://doi.org/10.1016/j.commatsci.2025.113811},
url = {https://www.sciencedirect.com/science/article/pii/S0927025625001545},
author = {Wenwen Zou and Juan Zhang and Jie Xu and Kai Jiang},
}

@article{ShechtmanMet,
  title = {Metallic Phase with Long-Range Orientational Order and No Translational Symmetry},
  author = {Shechtman, D. and Blech, I. and Gratias, D. and Cahn, J. W.},
  journal = {Phys. Rev. Lett.},
  volume = {53},
  issue = {20},
  pages = {1951--1953},
  numpages = {0},
  year = {1984},
  month = {Nov},
  publisher = {American Physical Society},
  doi = {10.1103/PhysRevLett.53.1951},
  url = {https://link.aps.org/doi/10.1103/PhysRevLett.53.1951}
}

@article{jiang2024numerical,
  title={Numerical methods and analysis of computing quasiperiodic systems},
  author={Jiang, Kai and Li, Shifeng and Zhang, Pingwen},
  journal={SIAM J. Numer. Anal.},
  volume={62},
  number={1},
  pages={353--375},
  year={2024},
  publisher={SIAM}
}

@article{fan2025representation,
  title={Representation of quasi-periodic functions and {H}ausdorff-{Y}oung inequalities for {B}esicovitch almost periodic functions},
  author={Fan, Aihua and Jiang, Kai and Zhang, Pingwen},
  journal={arXiv preprint arXiv:2512.06821},
  year={2025}
}

@article{scheiber2016ab,
  title={Ab initio calculations of grain boundaries in bcc metals},
  author={Scheiber, Daniel and Pippan, Reinhard and Puschnig, Peter and Romaner, Lorenz},
  journal={Model. Simul. Mater. Sci. Eng.},
  volume={24},
  number={3},
  pages={035013},
  year={2016},
  publisher={IOP Publishing}
}

@article{bollmann1982crystal,
  title={Crystal lattices, interfaces, matrices: an extension of crystallography},
  author={Bollmann, Walter},
  journal={Geneva},
  year={1982}
}

@article{bao2024convergence,
  title={Convergence Analysis for Bregman Iterations in Minimizing a Class of Landau Free Energy Functionals},
  author={Bao, Chenglong and Chen, Chang and Jiang, Kai and Qiu, Lingyun},
  journal={SIAM J. Numer. Anal.},
  volume={62},
  number={1},
  pages={476--499},
  year={2024},
  publisher={SIAM}
}

@article{levine1984quasicrystals,
  title={Quasicrystals: a new class of ordered structures},
  author={Levine, Dov and Steinhardt, Paul Joseph},
  journal={Phys. Rev. Lett.},
  volume={53},
  number={26},
  pages={2477},
  year={1984},
  publisher={APS}
}

@article{deymier1991experimental,
  title={Experimental evidence for a structural unit model of quasiperiodic grain boundaries in aluminum},
  author={Deymier, PA and Shamsuzzoha, M and Weinberg, JD},
  journal={J. Mater. Res.},
  volume={6},
  number={7},
  pages={1461--1468},
  year={1991},
  publisher={Cambridge University Press}
}

@article{dahmen2010correlation,
  title={Correlation between Atomic Structure and Superglide of an Incommensurate Grain Boundary in Au},
  author={Dahmen, U and Radetic, T and Ye, J and Minor, AM and Caliste, D and Lancon, F},
  journal={Microscopy and Microanalysis},
  volume={16},
  number={S2},
  pages={1442--1443},
  year={2010},
  publisher={Cambridge University Press}
}

@article{sutton1988irrational,
	title = {Irrational tilt grain boundaries as one-dimensional quasicrystals},
	journal = {Acta Metall.},
	volume = {36},
	number = {5},
	pages = {1291-1299},
	year = {1988},
	issn = {0001-6160},
	author = {A.P. Sutton}
}

@article{jiang2022tilt,
  title={Tilt grain boundaries of hexagonal structures: a spectral viewpoint},
  author={Jiang, Kai and Si, Wei and Xu, Jie},
  journal={SIAM J. Appl. Math.},
  volume={82},
  number={4},
  pages={1267--1286},
  year={2022},
  publisher={SIAM}
}

@article{li2019atomistic,
  title={Atomistic simulations of energies for arbitrary grain boundaries. {P}art {I}: Model and validation},
  author={Li, Saiyi and Yang, Liang and Lai, Chunming},
  journal={Comput. Mater. Sci.},
  volume={161},
  pages={330--338},
  year={2019},
  publisher={Elsevier}
}

@article{ranganathan1966geometry,
  title={On the geometry of coincidence-site lattices},
  author={Ranganathan, S},
  journal={Acta Crystallogr.},
  volume={21},
  number={2},
  pages={197--199},
  year={1966},
  publisher={International Union of Crystallography}
}

@article{masakova2000substitution,
  title={Substitution rules for aperiodic sequences of the cut and project type},
  author={Mas{\'a}kov{\'a}, Zuzana and Patera, Jir{\'\i} and Pelantov{\'a}, Edita},
  journal={J. Phys. A: Math. Gen.},
  volume={33},
  number={48},
  pages={8867},
  year={2000},
  publisher={IOP Publishing}
}

@incollection{demkowicz2008interfaces,
  title={Interfaces between dissimilar crystalline solids},
  author={Demkowicz, Michael J and Wang, Jian and Hoagland, Richard G},
  booktitle={Dislocations in solids},
  volume={14},
  pages={141--205},
  year={2008},
  publisher={Elsevier}
}

@article{romeu2003interfaces,
  title={Interfaces and quasicrystals as competing crystal lattices: Towards a crystallographic theory of interfaces},
  author={Romeu, David},
  journal={Phys. Rev. B},
  volume={67},
  number={2},
  pages={024202},
  year={2003},
  publisher={APS}
}

@article{sutton1983structure,
  title={On the structure of tilt grain boundaries in cubic metals {I}. {S}ymmetrical tilt boundaries},
  author={Sutton, Adrian Peter and Vitek, V},
  journal={Philos. Trans. R. Soc. A},
  volume={309},
  number={1506},
  pages={1--36},
  year={1983},
  publisher={The Royal Society London}
}

@article{gratias1988hidden,
  title={Hidden symmetries in general grain boundaries},
  author={Gratias, D and Thalal, A},
  journal={Philos. Mag. Lett.},
  volume={57},
  number={2},
  pages={63--68},
  year={1988},
  publisher={Taylor \& Francis}
}

@article{rivier1988quasicrystals,
  title={Quasicrystals at grain boundaries},
  author={Rivier, N and Lawrence, AJA},
  journal={Physica B+C},
  volume={150},
  number={1-2},
  pages={190--202},
  year={1988},
  publisher={Elsevier}
}

@article{sutton1992irrational,
  title={Irrational interfaces},
  author={Sutton, AP},
  journal={Prog. Mater. Sci.},
  volume={36},
  pages={167--202},
  year={1992},
  publisher={Elsevier}
}

@article{jiang2020efficient,
  title={Efficient numerical methods for computing the stationary states of phase field crystal models},
  author={Jiang, Kai and Si, Wei and Chen, Chang and Bao, Chenglong},
  journal={SIAM J. Sci. Comput.},
  volume={42},
  number={6},
  pages={B1350--B1377},
  year={2020},
  publisher={SIAM}
}

@article{jiang2025approximation,
  title={On the approximation of quasiperiodic functions with {D}iophantine frequencies by periodic functions},
  author={Jiang, Kai and Li, Shifeng and Zhang, Pingwen},
  journal={SIAM J. Math. Anal.},
  volume={57},
  number={1},
  pages={951--978},
  year={2025},
  publisher={SIAM}
}

@online{GBsoftware,
    author={Jiang, Kai and Si, Wei},
    title={QuasiPeriodic Interface Simulation Platform 2.0},
    howpublished={National Copyright Administration of the People’s Republic of China},
    year={2025},
    note={2025SR0122883, \url{https://github.com/KaiJiangMath/interface-qp-mpi}}
}

@article{shamsuzzoha1991atomic,
  title={The atomic structure of a [100], {$45^\circ$} twist plus {$17.5^\circ$} tilt grain boundary in aluminium by high-resolution electron microscopy},
  author={Shamsuzzoha, M and Smith, David J and Deymier, Pierre A},
  journal={Philos. Mag. A},
  volume={64},
  number={3},
  pages={719--733},
  year={1991},
  publisher={Taylor \& Francis}
}

@article{shamsuzzoha1999structural,
  title={Structural Study of a [100] {$45^\circ$} Twist Plus {$7.5^\circ$} Tilt Grain Boundary in Aluminium by {HREM}},
  author={Shamsuzzoha, M and Deymier, PA and Smith, David J},
  journal={Mater. Res. Symp. Proc.},
  volume={589},
  pages={353},
  year={1999},
  publisher={Cambridge University Press}
}

@article{shamsuzzoha2013geometrical,
  title={Geometrical Construction of {$\langle UVW \rangle$} {$(\Sigma/\Sigma_n)_{m\pm 1}$}-{$90^\circ$} Twist Quasi-Periodic Bi-Crystals and Their Quasi-Periodic Grain Boundaries in Cubic Crystals},
  author={Shamsuzzoha, M},
  journal={TMS 2013 Suppl. Proc.},
  pages={737--744},
  year={2013},
  publisher={John Wiley \& Sons, Inc. Hoboken, NJ, USA}
}

@inproceedings{penisson1998high,
  title={High resolution study of a quasiperiodic grain boundary in gold},
  author={P{\'e}nisson, Jean-Michel and Lan{\c{c}}on, F and Dahmen, U},
  booktitle={Mater. Sci. Forum},
  volume={294--296},
  pages={27--34},
  year={1998},
}

@article{gautam2013atomic,
  title={Atomic structure characterization of an incommensurate grain boundary},
  author={Gautam, A and Ophus, C and Lancon, F and Radmilovic, V and Dahmen, U},
  journal={Acta Mater.},
  volume={61},
  number={13},
  pages={5078--5086},
  year={2013},
  publisher={Elsevier}
}

@article{bowers2016step,
  title={Step coalescence by collective motion at an incommensurate grain boundary},
  author={Bowers, ML and Ophus, C and Gautam, A and Lancon, Frederic and Dahmen, Ulrich},
  journal={Phys. Rev. Lett.},
  volume={116},
  number={10},
  pages={106102},
  year={2016},
  publisher={APS}
}

@article{banadaki2018efficient,
  title={An efficient Monte Carlo algorithm for determining the minimum energy structures of metallic grain boundaries},
  author={Banadaki, Arash Dehghan and Tschopp, Mark A and Patala, Srikanth},
  journal={Comput. Mater. Sci.},
  volume={155},
  pages={466--475},
  year={2018},
  publisher={Elsevier}
}

@article{watanabe2011grain,
  title={Grain boundary engineering: historical perspective and future prospects},
  author={Watanabe, Tadao},
  journal={J. Mater. Sci.},
  volume={46},
  number={12},
  pages={4095--4115},
  year={2011},
  publisher={Springer}
}

@book{bollmann2012crystal,
  title={Crystal defects and crystalline interfaces},
  author={Bollmann, Walter},
  year={2012},
  publisher={Springer Science \& Business Media}
}

@article{sutton1989quasiperiodicity,
  title={Quasiperiodicity in irrational interfaces},
  author={Sutton, Adrian P},
  journal={Phase Transitions},
  volume={16},
  number={1-4},
  pages={563--574},
  year={1989},
  publisher={Taylor \& Francis}
}

@book{friedel1926leccons,
  title={Le{\c{c}}ons de cristallographie},
  author={Friedel, Georges},
  year={1926},
  publisher={Berger-Levrault}
}

\end{document}